\documentclass{JHEP3}
\usepackage{amsmath}
\usepackage{amssymb}
\usepackage{graphicx}
\newcommand{\be}{\begin{equation}}
\newcommand{\ee}{\end{equation}}
\newcommand{\bea}{\begin{eqnarray}}
\newcommand{\eea}{\end{eqnarray}}
\newcommand{\pd}{\partial}

\def\APJ{ Astroph.~J.~}

\title{Bouncing and Accelerating Solutions in Nonlocal Stringy Models}
\author{Irina~Ya.~Aref'eva\\
Steklov Mathematical Institute, Russian Academy of Sciences,\\
Gubkina st. 8, 119991, Moscow, Russia\\ E-mail:
\email{arefeva@mi.ras.ru}}
\author{Liudmila~V.~Joukovskaya\\
DAMTP, Centre for Mathematical Sciences, University of Cambridge,
Wilberforce Road, CB3 0WA, Cambridge, UK; \\  Steklov Mathematical
Institute, Russian Academy of Sciences, \\ Gubkina
st. 8, 119991, Moscow, Russia\\
E-mail: \email{l.joukovskaya@damtp.cam.ac.uk}}
\author{Sergey~Yu.~Vernov\\ Skobeltsyn
Institute of Nuclear Physics, Moscow State University, \\ Vorobyevy
Gory, 119991, Moscow, Russia\\ E-mail:
\email{svernov@theory.sinp.msu.ru}}


\abstract{A general class of cosmological models driven by a
nonlocal scalar field inspired by string field theories is
studied. In particular cases the scalar field is a string dilaton
or a string tachyon.
 A  distinguished  feature of these models
is  a crossing of the phantom divide. We  reveal the nature of
this phenomena showing that it is caused by an equivalence of the
initial nonlocal model to a model with an infinite number of local
fields some of which are ghosts. Deformations of the model that
admit exact solutions are constructed. These deformations contain
locking potentials that stabilize solutions. Bouncing and
accelerating solutions are presented.}

\preprint{hep-th/0701184}

\begin{document}

\section{Introduction}
 Field theories which violate the null energy
condition (NEC) \cite{Hawking-Ellis,Venziano} are of  interest for
the solution of the cosmological singularity problem and for the
construction of cosmological dark energy models with the state
parameter $w<-1$.

One of the first attempts to apply string theory to cosmology \cite{string-cosmo} was
related to the problem of the cosmological singularity \cite{Venziano}. A possible way to
avoid cosmological singularities consists of dealing with nonsingular bouncing
cosmological solutions. In these scenarios the Universe contracts before the bounce
\cite{Turok}. Such models have strong coupling and higher-order string
corrections are inevitable. It is important to construct nonsingular
bouncing cosmological solutions in order to make a concrete prediction
of bouncing cosmology.

Present cosmological observations \cite{cosmo-obser} do not exclude
an evolving dark energy (DE) state parameter $w$, whose current
value is less than  $-1$, that means the violation of the NEC (see
\cite{DE1,DE2} for a review of DE problems and \cite{0312430} for a
search  for a super-acceleration phase of the Universe).

A simple possibility to violate NEC is just to deal with a phantom
field. The phantom field is  unstable. There are
general arguments that coupled scalar-gravity models violating the
NEC are unstable (\cite{0301273}-\cite{0606091} and refs. therein).
 At the same time  a phantom  model could be an
approximation to a nonlocal model  that has no problems with
instability \cite{IA1}.  A simple example of such a model is a
model with a scalar nonlocal action $(e^{-\square_g}\phi)^2.$ In
the second order derivative approximation $e^{-\square_g}\approx
1-\square_g$  this model is equivalent to a phantom
 one but
  does not have  problems with instability.
This type of models does appear in String Field Theory (SFT)(see
\cite{review-sft} for a review) and in the p-adic string models
\cite{p-adic}. The model with the particular kinetic term
mentioned above is in fact realized in the p-adic string near a
perturbative vacuum and is expected to be realized in the Vacuum
String Field Theory (VSFT)~\cite{VSFT}.

The purpose of this paper is the study of this type of models. As a
model we consider a SFT inspired nonlocal dilaton action.
Distinguished features of the model are the invariance of the action
under the shift of the dilaton field to a constant as well as a
presence of infinite number of higher derivatives terms. A more
general family of nonlocal models loosing the invariance of the
nonlocal dilaton is also considered. For special values of the
parameters the models describe  linear approximations to the cubic
bosonic or nonBPS fermionic SFT nonlocal tachyon models, or p-adic
string models \cite{IA1}, \cite{AJ}-\cite{K}. The NonBPS fermionic
string field  tachyon nonlocal model has been considered as a
candidate for the dark energy~\cite{IA1}.  Several string-inspired
and braneworld dark energy models have been recently proposed (see
for example \cite{0605701}-\cite{0605039} and refs. therein). About
a study of the tachyon dynamics with the Born-Infeld action see
\cite{DBIreview,BSFT,Taylor}.

We discuss a possibility to stabilize the model that violates the
NEC in the  flat space-time  at the cost of  adding extra
interaction terms in the Friedmann background. One of the lessons
from a study nonlocal dynamics in the flat case is a sensitivity
of the stability problem to the form of the interaction term
\cite{Sen-g,ZW,AJK,LB,Vladimirov,AJ,Trento}. We use the
Weierstrass product to present the nonlocal field in terms of an
infinite number of local fields~\cite{AV}. Some of these local
fields are ghosts, which violate the NEC and are unstable. The
model is linear and admits exact solutions in the flat space-time.
In non-flat case we get the same exact solutions after a
deformation of the model. We used a similar approach to construct
effective SFT inspired phantom models~\cite{AKV,AJ,AKVtwofields}.

Another  recently proposed model which violates the
NEC and has higher derivatives is the ghost-condensation model
\cite{ghost-condensation}.
%It is discussed in \cite{AV}.
Vector-scalar and tensor-scalar models that  violate NEC
and are stable in
some region have been proposed in \cite{Rubakov,Starobinsky}, respectively.

The paper is organized as follows. In Section 2 we describe our
strategy to the study of stringy inspired models. In Section 3 we
present general solutions   of the models in the flat case. Then
we use some approximation to study these dynamics in the Friedmann
metric and discuss cosmological properties of the constructed
solutions.

\section{Set up}

 In this paper we consider a model of gravity coupling with a nonlocal scalar field
 which induced by strings field theory
 \begin{equation}
\label{ACTION} S=\int d^4x\sqrt{-g}\left(\frac{M_p^2}{2}R+
\frac{M_s^4}{g_4}\left(\frac{1}{2}\phi\,F(-\square_g/M_s^2)\phi
-\Lambda^\prime \right)\right),
\end{equation}
where $g$ is the metric,
$\square_g=\frac1{\sqrt{-g}}\pd_{\mu}\sqrt{-g}g^{\mu\nu}\pd_{\nu}$,
$M_p$
 is  a mass Planck, $M_s$ is a characteristic string scale
 related with the  string tension $ \alpha^{\prime}$,
$M_s=1/\sqrt{ \alpha^{\prime}}$, $\phi$ is a dimensionless scalar
field (tachyon or dilaton), $g_4$ is a dimensionless four
dimensional effective coupling constant related with the ten
dimensional  string coupling constant $g_0$ and the compactification
scale. $\Lambda=\frac{M_s^4}{g_4}\Lambda^\prime $ is an effective
four dimensional cosmological constant.

The form of the function $F$ is inspired
by a nonlocal action appeared in   string field theories.
In   particular cases
\begin{equation}
\label{F} F(z)=-\xi^2 z+1-ce^{-2z},
\end{equation}
$\xi$ is a real
parameter and $c$ is a positive constant.
Using dimensional space-time variables
and after a rescaling
we can rewrite
(\ref{ACTION}) for $F$ given by (\ref{F}) as
follows
\begin{equation}
\label{action-1}  \int
d^4x\sqrt{-g}\left(\frac{m_p^2}{2}R+\frac{\xi^2 }{2}
\phi\,\square_g\phi+
 \frac{1}{2}(\phi^2-c\Phi^2)-\Lambda^\prime\right),
\end{equation}
where $\Phi=e^{\square_g}\phi$ and $m_p^2=g_4M_p^2/M_s^2$.
Generally speaking the string scale does not coincide with the
Plank mass~\cite{IA_marion,Cline}. This gives a possibility to get
a realistic value of $\Lambda$~\cite{IA_marion}.

The form of the term $(e^{\square_g}\phi)^2$
 is analogous to the form of the
interaction in the action for the string field tachyon in non-flat
background \cite{IA1}, which is a generalization of the SFT
tachyon interaction term   in a flat background
\cite{KS,West,review-sft,ZW,AJK}. At some particular values of
$\xi^2$ and $c$ this action appears in a linear approximation to
SFT actions \cite{Witten}-\cite{BSZ} and in a non-flat background
has been considered in \cite{IA1,AJ,Calcagni,AK,K}. The case of
the open Cubic Superstring Field Theory (CSSFT)~\cite{IA1,AK}
tachyon corresponds to
$\xi^2={}-1/\left(4\ln\left(\frac{4}{3\sqrt{3}}\right)\right)\approx
0.9556$ and $c=3$. We consider in detail action (\ref{action-1})
at $c=1$, which is invariant under translation $\phi \to \phi
+const$.

We take the metric in the form
\begin{equation}
\label{mFr} ds^2={}-dt^2+a^2(t)\left(dx_1^2+dx_2^2+dx_3^2\right)
\end{equation}
and   get the following equation of motion for the space homogeneous scalar field $\phi$:
\begin{equation}
F(-{\cal D})\phi =0, \label{eom-phi-H}
\end{equation}
where
\begin{equation}
\label{cal-D} {\cal D}\equiv {}- \partial_t^2- 3H(t)\partial_t,
\quad  H=\frac{\dot{a}}{a} \quad {\mbox {and}} \quad \dot a\equiv
\partial_t a.
\end{equation}

The Friedmann equations have the following form
\begin{equation}
\begin{split}
3H^2&=\frac{1}{m_p^2}~{\cal E},
\\3H^2+2\dot H&={}-\frac{1}{m_p^2}~{\cal P},
\end{split}
\label{eomprho}
\end{equation}
where the energy  and the pressure are  obtained from  the action
(\ref{ACTION}) using standard formula
\begin{equation}\label{Tmunu}
T_{\mu\nu}=-\frac{2}{\sqrt{-g}}\frac{\delta{S}}{\delta
g^{\mu\nu}},\qquad T_{\mu\nu}=\mbox{diag}\{{\cal E},{\cal P},{\cal
P},{\cal P}\}. \end{equation} For the case of $F$ given by
(\ref{F}) the energy  and the pressure have additional nonlocal
terms ${\cal E}_{nl1}$ and ${\cal E}_{nl2}$ \cite{AJK,Yang,LJ}
\begin{equation}
\label{E-phi-H-p}
\begin{split}
 {\cal E}&={\cal E}_k+{\cal E}_p+{\cal E}_{nl2}+{\cal E}_{nl1}+\Lambda^\prime,\\
{\cal P}&={\cal E}_k-{\cal E}_p+{\cal E}_{nl2}-{\cal E}_{nl1}-\Lambda^\prime.
\end{split}
\end{equation}
Nonlocal term ${\cal E}_{nl1}$ plays a role of an extra potential
term and ${\cal E}_{nl12}$ a role of an extra kinetic term. The
explicit form of the terms in the R.H.S. of (\ref{E-phi-H-p}) is
\begin{equation}
\label{El-Enl}
\begin{split}
 {\cal E}_k&=\frac{\xi^2}{2}(\partial\phi)^2,
 \qquad {\cal E}_p={}-\frac{1}{2}\left(\phi^2-c(e^{\cal
D}\phi)^{2}\right),\\
{\cal E}_{nl1}&=c\int_{0}^{1}\left(e^{(1+\rho) {\cal D}} \phi\right)
 \left({}-{\cal D} e^{(1-\rho){\cal D}}  \phi\right) d \rho,
\\{\cal
E}_{nl2}&=-c\int_{0}^{1}\left(\partial e^{(1+\rho){\cal D}}
\phi\right) \left(\partial e^{(1-\rho) {\cal D}}\phi\right) d \rho.
\end{split}
\end{equation}
\par
$~$
\par

Our strategy is the following:

First, we study the dynamics of the model (\ref{action-1}) in the flat case.
\begin{itemize}
\item We show that solutions have the form of plane waves. There
are special values of parameters for which plane waves should be
multiplied on linear functions.  We calculate the energy and
pressure on the corresponding solutions.

\item We present  dynamics of the nonlocal model in terms of an
infinite number of local fields \cite{AV} and show that the energy
and pressure densities of the nonlocal model are reproduced by the
energy and pressure densities of the corresponding local models. For
this purpose we use the Weierstrass product representation for the
function $F$ in (\ref{ACTION}),
\begin{equation}
F(z)=e^{f(z)}\prod_n \left(1-\frac{z}{\alpha^2_n}\right),
\label{W-z}
\end{equation}
where $\alpha^2_n$ are complex numbers, and represent the flat
analog of (\ref{ACTION}) as
\begin{equation}
S_{flat}=\frac12\int d^4x \phi F(-\square )\phi \sim \frac12 \sum
\left[\epsilon _n \psi_n e^{f(-\square)}(\square+\alpha_n^2)\psi_n
+c.c.\right],
\end{equation}
where $\square$ is the d'Alembertian in the flat space-time.

\item We consider approximated models obtained
by a truncation of number of local fields.
\end{itemize}

Then we consider the Friedmann Universe. There are two ways to
study dynamics in the Friedmann metric:
\begin{itemize}

\item  One can use the found expressions for the energy and pressure
in the flat case, $E(\phi_0)$ and $P(\phi_0)$,  to calculate the
corresponding Hubble parameter $H_0$, then using this Hubble
parameter calculate a perturbation of the flat solution of equation
of motion and so on:
\begin{equation}\phi= \phi_0+ \phi_1(H_0)+...,~~~~~
H=H_0(\phi_0)+H_1(\phi_0,\phi_1)+...
\end{equation}
\item One can search for
 deformations of the model that admit the same exact solutions as in
 the flat
case and try to argue  that the deformed models describe the initial
model with a good accuracy.
\end{itemize}

Both ways permit to find the first approximations to the models
(\ref{eom-phi-H}), (\ref{eomprho}). The first way have been used
in~\cite{AK}. In this paper we will follow the second way.

To this goal we use a representation of nonlocal  dynamics given
by action (\ref{ACTION}) in terms of local fields
\begin{equation}
\int  d^4x\sqrt{-g}\left(\frac{m_p^2}{2}R+\frac12  \sum
\left[\epsilon _n \psi_n
e^{f(-\square_g)}(\square_g+\alpha_n^2)\psi_n+\Lambda^\prime
+c.c.\right]\right). \label{g-inf-pairs}
\end{equation}
We perform a deformation of this model by several steps. First,  we consider an approximation to the model
(\ref{g-inf-pairs})
in the form
\begin{equation}
 \int  d^4x
\sqrt{-g}\left(\frac{m_p^2}{2}R+ \sum \left[\frac{\epsilon _n}{2}
\psi_n e^{f(\alpha_n^2)}(\square_g+\alpha_n^2)\psi_n+\Lambda^\prime
+c.c.\right]\right)
\end{equation}
Second, we restrict a number of local fields and, third, we add
potentials of the order $1/m_p^2$  in which  $\Lambda^\prime$ is
also included:
\begin{equation}
\label{def-pot}
 \int  d^4x
\sqrt{-g}\left(\frac{m_p^2}{2}R+ \sum \left[\frac{\epsilon _n}{2}
\psi_n e^{f(\alpha_n^2)}(\square_g+\alpha_n^2)\psi_n
+c.c.\right]-{\cal V}(\psi_1,\dots,\psi_n)\right).
\end{equation}
such that solutions of
the field equations in the non-flat case are the same as the flat case.

Finally, we find the corresponding scale factor $a(t)$ and study cosmological
properties of  approximated solutions to our model.

\section{Flat Dynamics}

\subsection{General Solutions}
\subsubsection{Roots of the Characteristic Equation}
In the flat case  the action (\ref{ACTION}) has the following
form:
\begin{equation}
\label{action-1flat} S_{flat}=\frac{1}{2}\int d^4x \phi
F(-\square)\phi.
\end{equation}
Equation of motion on the space-homogeneous configurations
(\ref{eom-phi-H}) is reduced to the following linear equation:
\begin{equation}\label{F-eq} F(\partial ^2)\phi=0. \end{equation}A plane
wave
\begin{equation}
\phi=e^{\alpha t} \label{dec}
\end{equation}
is a solution of (\ref{F-eq})
if $\alpha$ is a root of the characteristic equation
\begin{equation}
F(\alpha^2)=0. \label{ch-F}
\end{equation}

For a case of $F$ given by (\ref{F}) equation (\ref{F-eq}) has the following form
\begin{equation}
-\xi^2\partial ^2 \phi + \phi- ce^{-2\partial^2}\phi=0 \label{1c}.
\end{equation}
This equation has an infinite number of derivatives and can be
treated as a pseudodifferential as well as an integral equation
\cite{Vladimirov}.

The corresponding characteristic equation:
\begin{equation}
F(\alpha^2)\equiv-\xi^2\alpha^2  + 1 - ce^{-2\alpha^2}=0 \label{5c}
\end{equation}
has the following solutions
\begin{equation}
\label{sol-lamb-k} \alpha_n={}\pm\frac{1}{2\xi}\sqrt{4+
2\xi^2 W_n\left({}-\frac{2ce^{-2/\xi^2}}{\xi^2}\right)}, \quad
n=0,\pm 1,\pm 2,...
\end{equation}
where $W_n$ is  the n-s branch of the  Lambert function satisfying a relation
$W(z)e^{W(z)}=z$. The
Lambert function is a multivalued function, so eq.~(\ref{5c}) has an
infinite number of roots.

Parameters $\xi$ and $c$ are real,
therefore if $\alpha_n$ is a root of (\ref{5c}), then the adjoined
number $\alpha_n^*$ is a root as well. Note that  if $\alpha_n$ is a
root of (\ref{5c}), then $-\alpha_n$ is a root too. In other words,
equation (\ref{5c}) has quadruples of complex roots
\begin{equation}
\alpha_{n,\pm\pm}= \pm \mathrm{Re}(\alpha_{n})\pm
i\mathrm{Im}(\alpha_{n}). \label{5c_pm}
\end{equation}

 If $\alpha^2=\alpha^2_0$ is a multiple root, then at this point
$F(\alpha^2_0)=0$ and $F'(\alpha^2_0)=0$. These equations give
that
\begin{equation}
\label{z-0} \alpha _0^2=\frac{1}{\xi^2}-\frac12,
\end{equation}
 hence $\alpha^2_0$ is a real number and all multiple roots of
 $F(\alpha^2_0)=0$ are either real or pure imaginary.
The multiple roots exist if and only if
\begin{equation}
\label{c-xi} c=\frac{\xi^2}{2e}e^{2/\xi^2}.
\end{equation}

Real roots for any $\xi$ and $c$, except $\xi^2=0$ and $c=\infty$,
are no more then double degenerated, because  $F''(\alpha^2_0)\neq
0$.

Summing up we note that according as the values of parameters $c$
and $\xi^2$ there exist the following types of the general real
solution of (\ref{1c}):
\begin{itemize}
\item If $c\neq\frac{\xi^2}{2e}e^{2/\xi^2}$ and  $c\neq 1$ then
the general real solution is

\begin{equation}
\label{phicomplex}
    \phi=\sum_{n} R_ne^{m_n t}+\sum_{n}\left(C_ne^{\alpha_n t}+{C_n^*}e^{\alpha_n^* t}\right),
\end{equation}
where $R_n$ and $C_n$ are arbitrary real and complex numbers
respectively.

\item If
$c=\frac{\xi^2}{2e}e^{2/\xi^2}>1$, then  to get the general real
solution one has to add to (\ref{phicomplex})
\bea
\label{deg}
    \phi_0=\tilde{R}_{1}te^{ m_0 t}+\tilde{R}_{2}te^{-m _0 t},
    \qquad m _0=\sqrt{\frac{1}{\xi^2}-\frac12}\quad
 &\mbox{if}&\xi^2<2,\\
 \phi_0=\tilde{C}_{1}te^{i\alpha _0 t}+\tilde{C}^*_{1}te^{-i\alpha _0 t},
 \qquad
\alpha_0=i\sqrt{\frac12-\frac{1}{\xi^2}}\quad &\mbox{if}&\xi^2>2.
\eea

\item
If $c=1$ then to get the general real solution one has to add to
(\ref{phicomplex})
\bea
\label{equsol}
    \phi_0=C_1t+C_0, \qquad &\mbox{if} \quad &\xi^2\neq 2,
\\
\label{equsol4}
    \phi_0=C_3t^3+C_2t^2+C_1t+C_0,\qquad &\mbox{if} \quad &\xi^2= 2.
\eea
\end{itemize}

\subsubsection{Real Roots of the Characteristic Equation}
 For some values of the parameters $\xi$ and $c$
eq.~(\ref{5c}) has real roots. To mark out real values of $\alpha$
we will denote real $\alpha$ as $m$: $m=\alpha$.

To determine values of the parameters at which eq.~(\ref{5c}) has real
roots we rewrite this equation in the following form:
\begin{equation}\label{equxi}
    \xi^2 = g(m^2,c), \quad {\mbox {where}} \quad
    g(m^2,c)=\frac{e^{2m^2}-c}{m^2e^{2m^2}}.
\end{equation}
The dependence of $g(m,c)$ on $m$ for different $c$ is presented in
Fig.~(\ref{xi2_m}). This function has a maximum at $m_{max}^2$
\begin{equation}
m_{max}^2=-\frac12-\frac12
W_{-1}\left(-\frac{e^{-1}}{c}\right), \label{max}
\end{equation}
provided $c$ is such that
$W_{-1}\left(-\frac{e^{-1}}{c}\right)<-1$, in the other words
$0<c<1$.

\FIGURE{
\includegraphics[width=45mm]{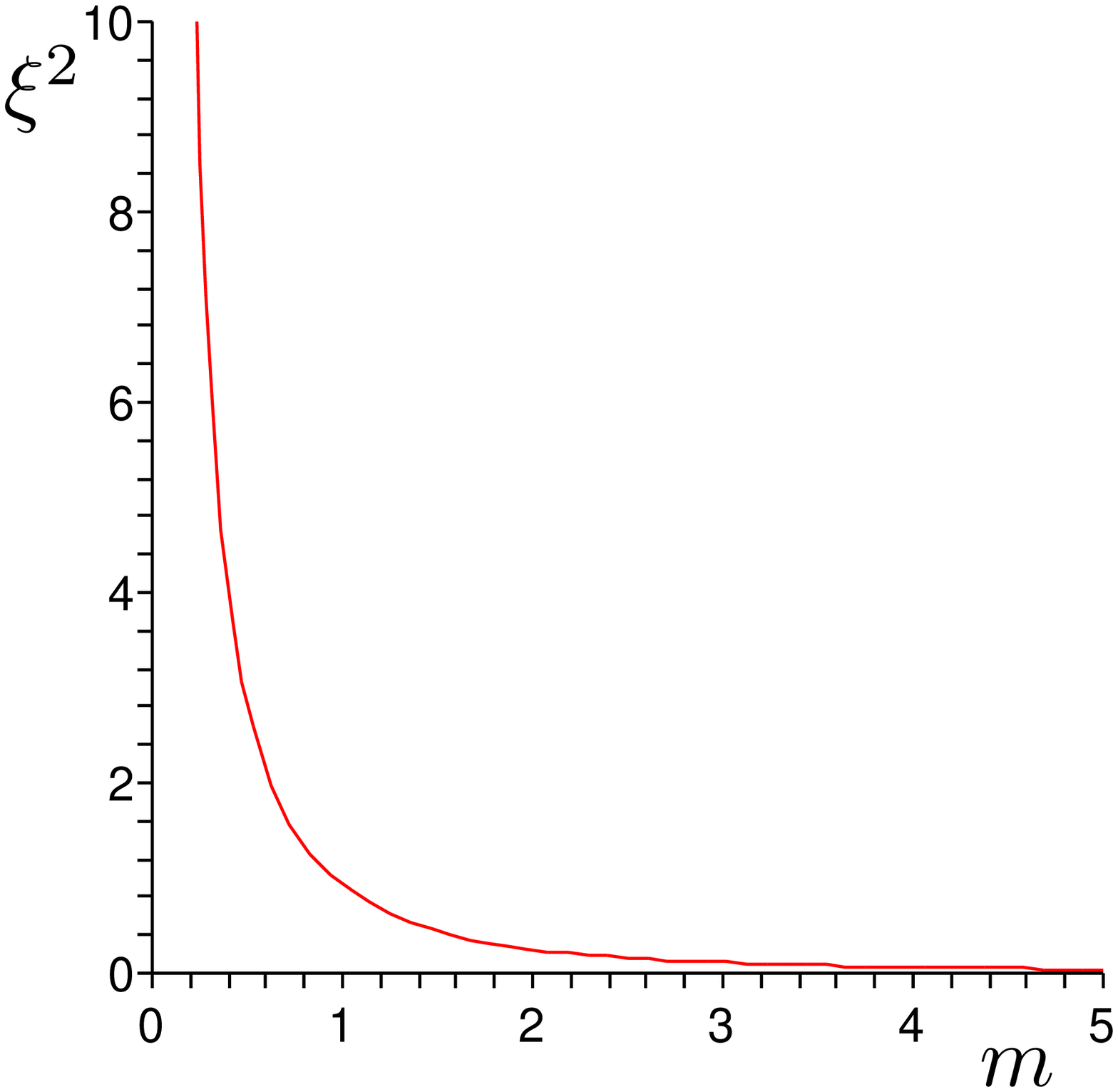} \ \ \ \ \
\includegraphics[width=45mm]{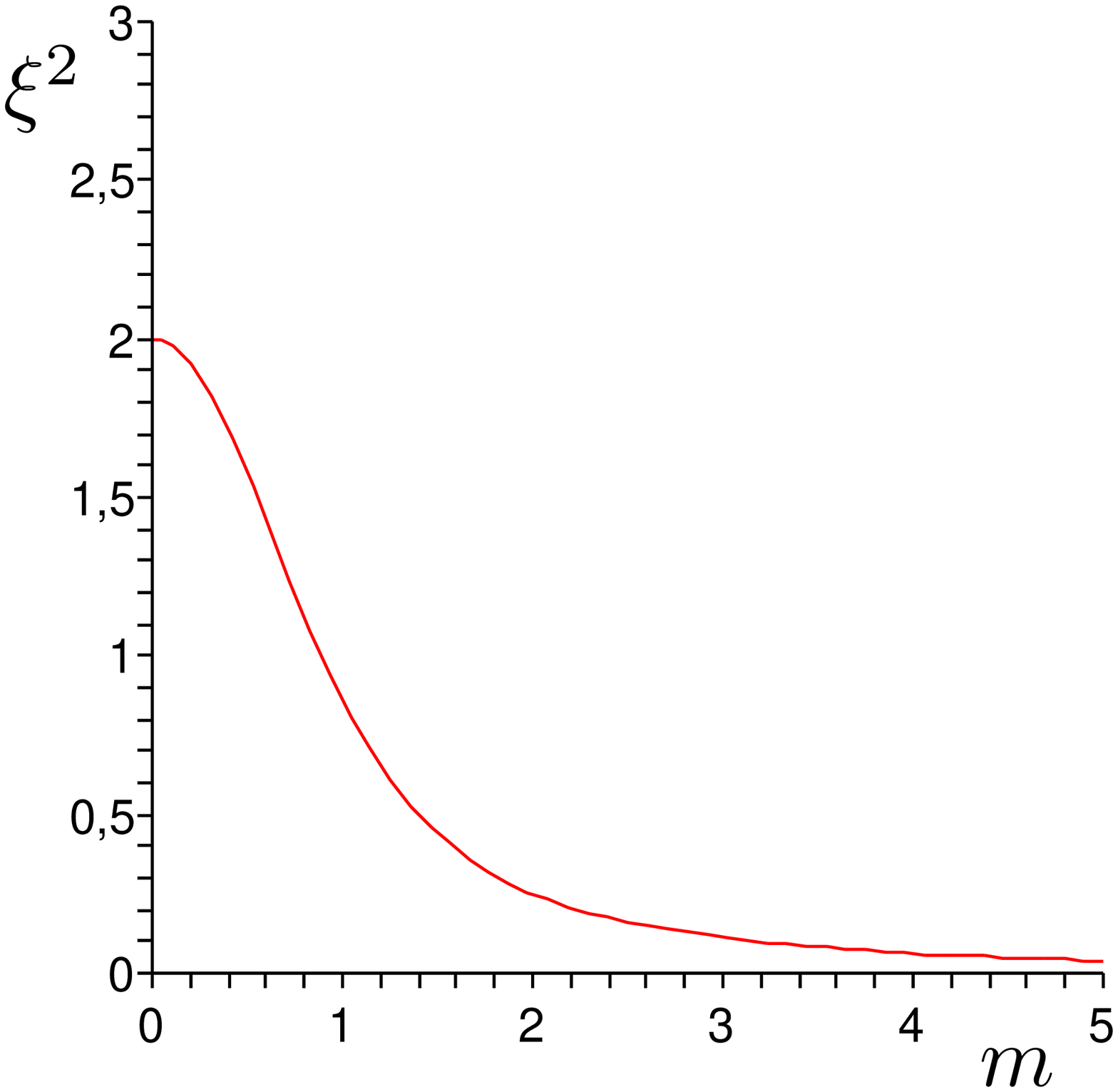} \ \ \ \ \
\includegraphics[width=45mm]{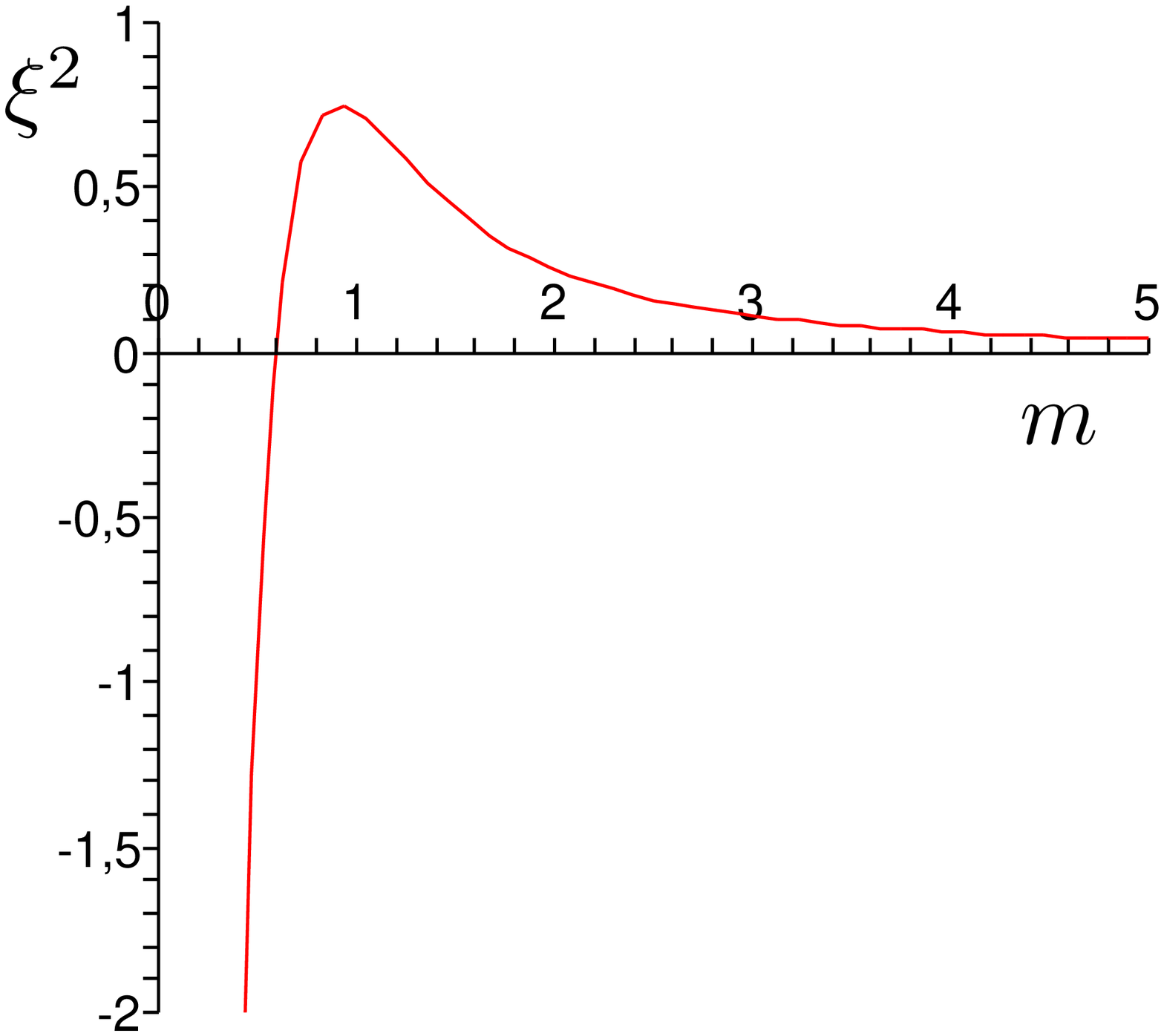}
\caption{ The dependence of  function $g(m^2,c)$, which is equal
to $\xi^2$, on $m$ at $c=1/2$ (left), $c=1$ (center) and $c=2$
(right). } \label{xi2_m}}

There are three different cases (see Fig.~\ref{xi2_m}).
\begin{itemize}
\item If $c<1$, then eq.~(\ref{5c}) has two simple real roots: $m=\pm
m_1$ for any values $\xi$.

\item If $c=1$, then eq.~(\ref{5c}) has a zero root. Nonzero real roots exist if and
only if $\xi^2>2$.

\item If $c>1$, then eq.~(\ref{5c}) has
\begin{itemize}
\item
no real roots for $\xi^2>\xi^2_{max}$, where
\begin{equation}
\label{xi-max} \xi_{max}^2=\frac{1-ce^{-2m_{max}^2}}{m_{max}^2}
={}-\frac{2}{W_{-1}(-e^{-1}/c)}
\end{equation}

\item two real double roots $m=\pm m_{max}$ for $\xi^2=\xi^2_{max}$

\item four real single roots for  $\xi^2<\xi^2_{max}$.
In this case we have the following restriction on real roots:
$m^2>\frac12\ln c$.
\end{itemize}
\end{itemize}

\subsubsection{Pure Imaginary Roots of the Characteristic Equation}
For some values of the parameters $\xi$ and $c$ eq.~(\ref{5c}) has
a pair of pure imaginary roots. Let us introduce a new real
variable $\mu=i\alpha$. From (\ref{1c}) we obtain
\begin{equation}
(\xi^2\mu^2  + 1)e^{-2\mu^2}=c. \label{5d}
\end{equation}

This equation is equivalent to
\begin{equation}
\xi^2 = \tilde{g}(\mu^2,c),\quad\mbox{where} \quad
\tilde{g}(\mu^2,c)=\frac{c-e^{-2\mu^2}}{\mu^2e^{-2\mu^2}}.
\label{5da}
\end{equation}
The dependence of $\tilde{g}(\mu^2,c)$ on $\mu$ for different $c$ is
presented in Fig.~\ref{xi2-mu}.

\FIGURE{ \centering
\includegraphics[width=45mm]{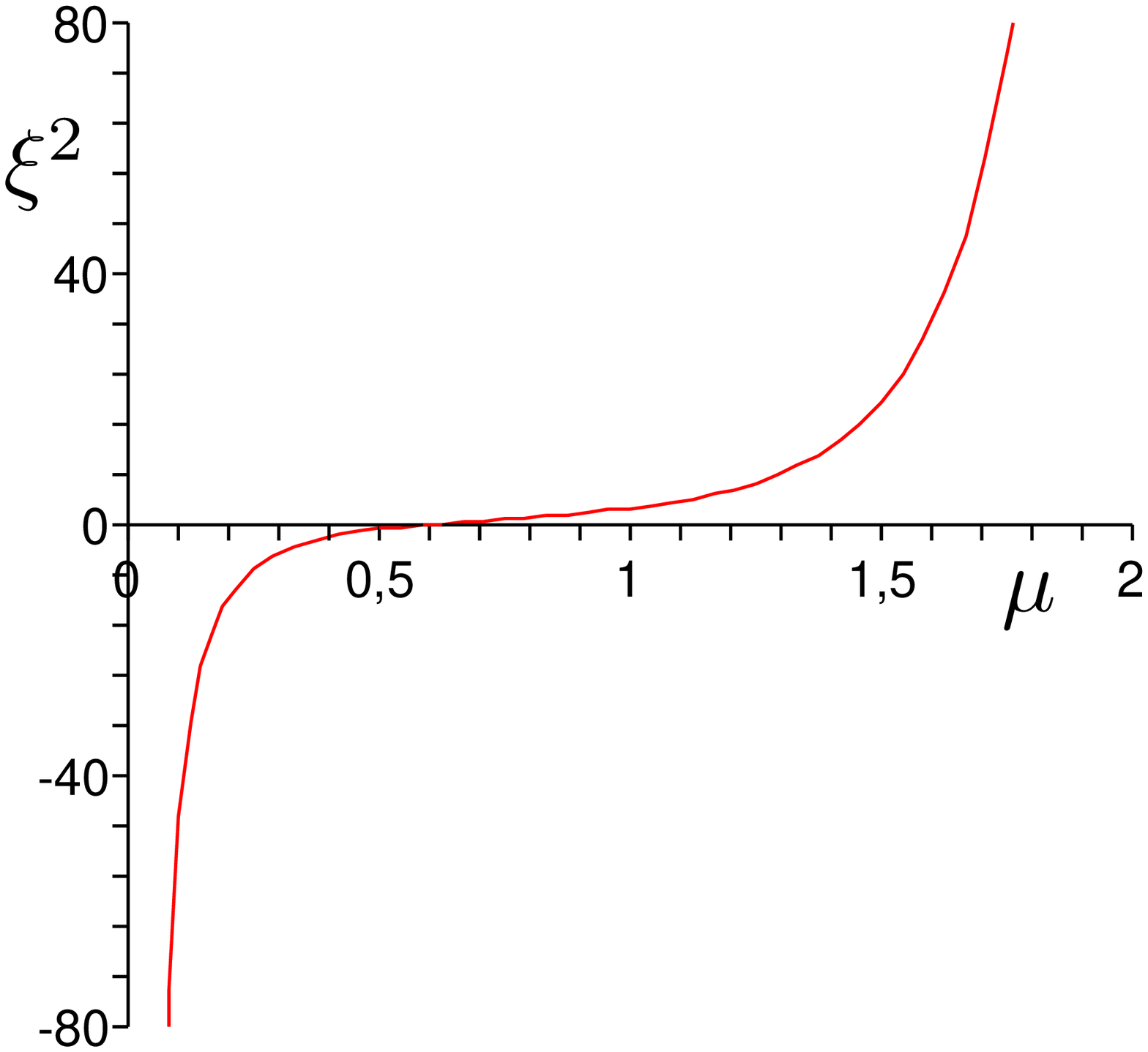} \ \ \ \ \
\includegraphics[width=45mm]{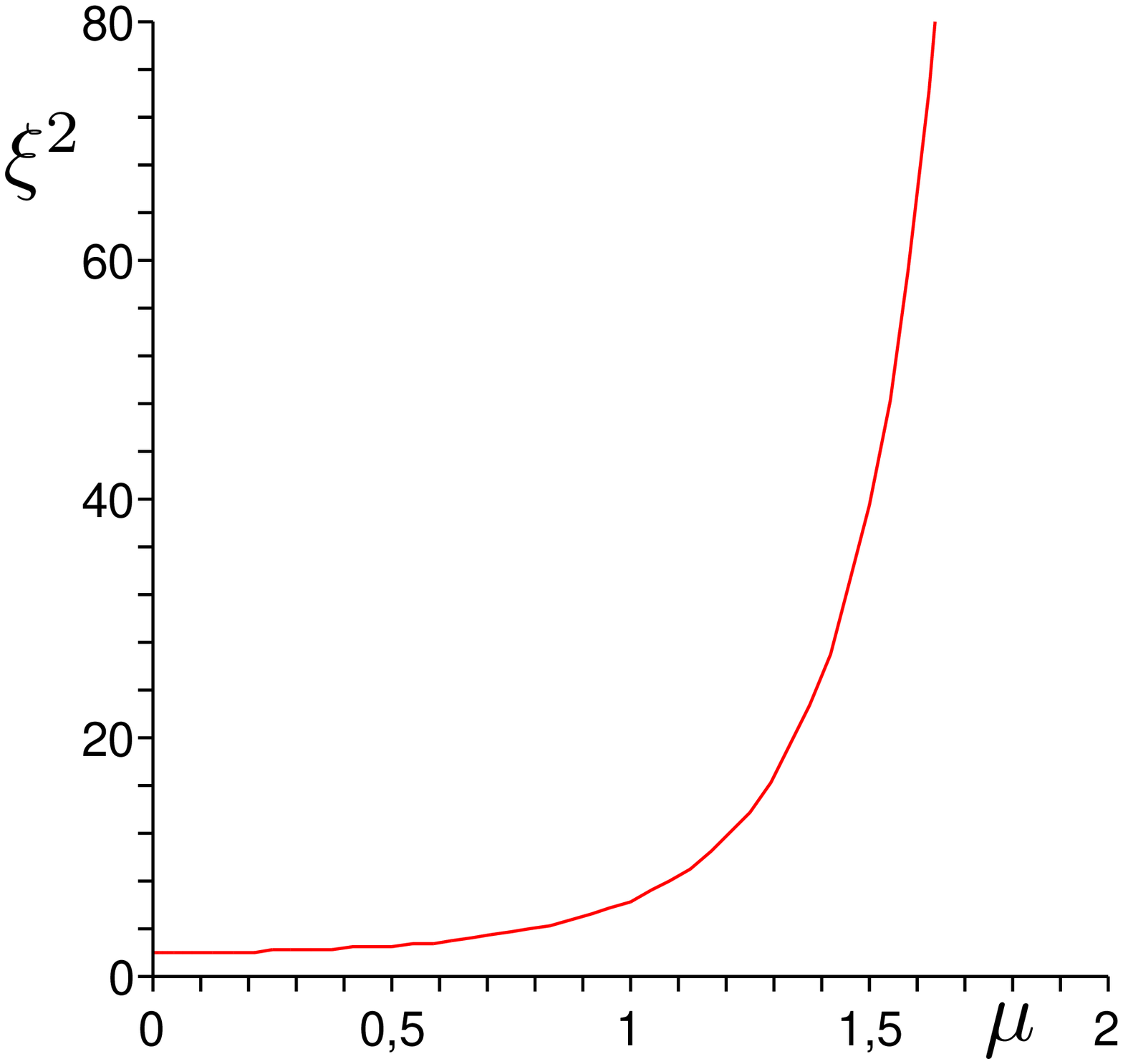} \ \ \ \ \
\includegraphics[width=45mm]{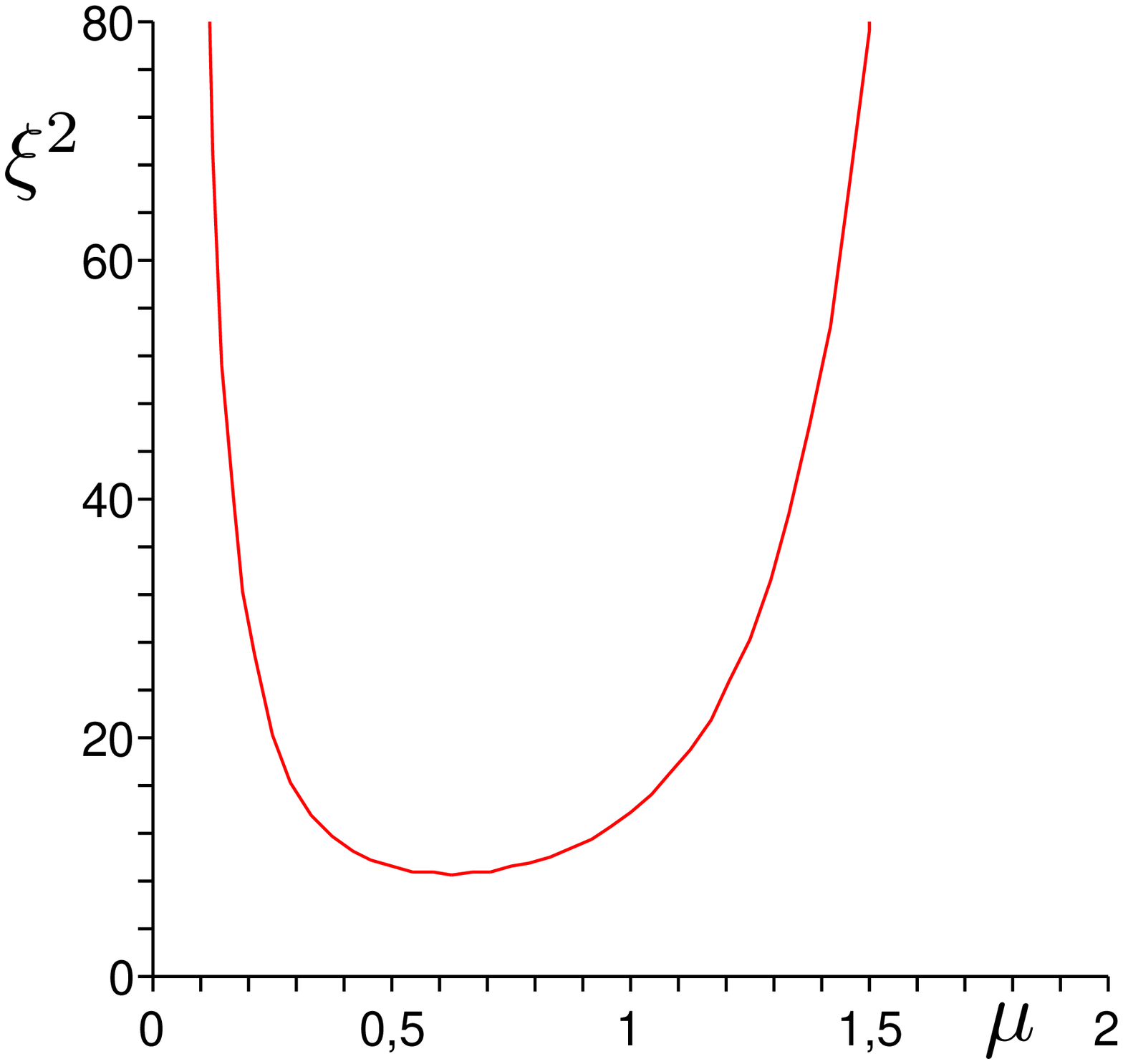}
\caption{ The dependence of  function $\tilde{g}(\mu^2,c)$, which
is equal to $\xi^2$,  on $\mu$ at $c=1/2$ (right), $c=1$ (center)
and $c=2$ (left). } \label{xi2-mu}}

 For different $c$ we have:
\begin{itemize}
\item For  $c<1$ there are two real simple roots $\mu=\pm\mu_1$,

\item For  $c=1$ nonzero real roots exist only if $\xi^2>2$,

\item For  $c>1$ real roots exist if and only if
\begin{equation}
\xi\geqslant
\xi_{min}={}-\frac{2}{W_{0}\left(-\frac{e^{-1}}{c}\right)}.
\end{equation}
 If $\xi=\xi_{min}$, then there exist two double real
roots: $\mu=\pm\mu_{min}$, where
\begin{equation}
\label{mu_min}
    \mu_{min}=\frac12\sqrt{2+2W_{0}\left(-\frac{e^{-1}}{c}\right)}.
\end{equation}
At $\xi>\xi_{min}$ eq.~(\ref{5d}) has four real simple roots.
\end{itemize}

\subsubsection{Roots at the SFT inspired values on $\xi^2$ and $c$}

Let us consider a special values of $\xi^2$ and $c$, which have
been obtain in the SFT inspired cosmological model~\cite{IA1,AK}.
From the action for the tachyon in the CSSFT~\cite{TCSFT} the
following equation has been obtained~\cite{AK}:
\begin{equation}
 (-\xi_0^2\tilde{\alpha}^2+1)=3e^{-\tilde{\alpha}^2/4},
\end{equation}
where
\begin{equation}
\xi_0^2={}-\frac{1}{4\ln\left(\frac{4}{3\sqrt{3}}\right)}\approx
0.9556.
\end{equation}

Substituting $\tilde{\alpha}=2\sqrt{2}\alpha$, we obtain
eq.~(\ref{5c}) with $\xi^2=8\xi_0^2$ and $c=3$. From (\ref{c-xi})
it is follows that all roots are simple. We obtain that
$\xi^2_{min}>\xi^2>\xi^2_{max}$, so there exist neither real roots
nor pure imaginary roots.

\subsection{Energy Density and Pressure}
\subsubsection{General Formula} Equation (\ref{1c}) has the conserved
energy (compare with \cite{Yang,LJ,AJ}), which is defined by the formula
that is a flat analog of (\ref{E-phi-H-p}). The energy density is
as follows:
\begin{equation}
\label{E} E=E_k+E_p+E_{nl1}+E_{nl2},
\end{equation}
where
\begin{equation}
\label{Ekp_flat} E_k=\frac{\xi^2}{2}(\partial\phi)^2, \qquad
E_p=-\frac{1}{2}\phi^2+\frac{c}{2}\Phi^{2},
\end{equation}
\begin{equation}
E_{nl1}=c\int_{0}^{1}\left(e^{-\rho
\partial^2} \Phi\right)
\left(\partial^2(e^{ \rho \partial^2}  \Phi)\right) d\rho, \qquad
 E_{nl2}=-c\int_{0}^{1}\left(\partial(e^{-\rho
\partial^2} \Phi)\right)\left(\partial(e^{ \rho \partial^2}  \Phi)\right)
d \rho.
\end{equation}

For the pressure
\begin{equation}
\label{p} P=E_k-E_p-E_{nl1}+E_{nl2},
\end{equation}
 we have  the following  explicit form
\begin{equation}
\label{p-exp}
P=\frac{\xi^2}{2}(\partial\phi)^2+\frac{1}{2}\phi^2-\frac{c}{2}\Phi^{2}
-c\int_{0}^{1}\left\{\left(e^{-\rho\partial^2} \Phi\right)
 \left(\partial^2 (e^{ \rho \partial^2} \Phi)\right) -\left(\partial(e^{-\rho
\partial^2}  \Phi)\right) \left(\partial(e^{ \rho\partial^2}
\Phi)\right)\right\} d \rho.
\end{equation}

Let us calculate  the energy density and pressure for the following
solution
\begin{equation}
\label{phiexpsum}
    \phi=\sum_{n=1}^{N} C_n e^{\alpha_n t},
\end{equation}
where $N$ is a natural number, $C_n$ are some constant and
$\alpha_n$ are solutions to eq. (\ref{5c}).

For $N=1$ and
\begin{equation}
\label{phione}\phi=Ce^{\alpha t}
\end{equation}
we obtain
\begin{eqnarray}
\label{E11-h} E(Ce^{\alpha t})&=&0,\\
\label{E12-h} P(Ce^{\alpha t})&=&C^2p_{\alpha}e^{2\alpha t}.
\end{eqnarray}
Hereafter we denote the energy density and pressure of function
$\phi(t)$ as the functionals $E(\phi)$ and $P(\phi)$, respectively,
and use the following notation
\begin{equation}
\label{pressure}
p_{\alpha}\equiv\alpha^2\left(\xi^2-2+2\xi^2\alpha^2\right).
\end{equation}

For $N=2$ and
\begin{equation}
    \phi=C_1e^{\alpha_1 t}+C_2e^{\alpha_2 t},\label{phi2e}
\end{equation}
where $\alpha_1$ and $\alpha_2$ are different roots of (\ref{5c})
 we have (see Appendix A for details)
\begin{equation}
E\left(C_1e^{\alpha_1 t}+C_2e^{\alpha_2 t}\right)=
-2C_1C_2p_{\alpha_1}, \qquad \mbox{at} \quad \alpha_2=-\alpha_1,
\end{equation}
 and
\begin{equation}
\label{Ediff}
    E\left(C_1e^{\alpha_1 t}+C_2e^{\alpha_2 t}\right)=0, \qquad\mbox{at} \quad
\alpha_2\neq{}-\alpha_1.
\end{equation}

The pressure $P(\phi)$ for solution (\ref{phi2e}) is
\begin{equation}
\label{52m}
  P\left(C_1e^{-\alpha t}+C_2e^{\alpha t}\right)=\left(C_1^2e^{-2\alpha
t}+C_2^2e^{2\alpha t}\right)p_\alpha.
\end{equation}

In the general case we have
\begin{equation}
\label{Egen}
    E\left(\sum_{n=1}^{N} C_n e^{\alpha_n t}\right)={}-2\sum_{n=1}^{N}\sum_{k=n+1}^N C_nC_k
p_{\alpha_n}\delta_{\alpha_n,-\alpha_k},
\end{equation}
where
\begin{equation}
\delta_{\alpha_n,-\alpha_k}=
\begin{cases} 1,&\alpha_n=-\alpha_k,\\0,&
\alpha_n\neq-\alpha_k.
\end{cases}
\end{equation}

 Note that all summands in (\ref{Egen}) are
integrals of motion, therefore, we explicitly show that
$E\left(\sum_{n=1}^{N} C_n e^{\alpha_n t}\right)$ is an integral of
motion. From formula (\ref{Egen}) we see that the energy density is
a sum of the crossing terms.
 At the same time the
pressure is a sum of "individual" pressures and has no crossing
term. In the case of an arbitrary finite number of summands the
pressure is as follows:
\begin{equation}
     P\left(\sum_{n=1}^{N} C_n e^{\alpha_n t}\right)=
     \sum_{n=1}^{N}C_n^2P\left(e^{\alpha_n t}\right)=\sum_{n=1}^{N}C_n^2
     p_{\alpha_n} e^{2\alpha_n t}.
\label{P-gen}
\end{equation}

If the parameters $\xi^2$ and $c$ are such that the characteristic
equation (\ref{5c}) have double roots, then eq.~(\ref{1c}) has the
following solution
\begin{equation}
\label{soldouble} \phi_0(t)=B_1 e^{\alpha_0 t}t+C_1e^{\alpha_0
t}+B_2e^{-\alpha_0 t}t+C_2e^{-\alpha_0 t},
\end{equation}
where $B_1$, $B_2$, $C_1$ and $C_2$ are constants, $\alpha_0\neq 0$
is defined by (\ref{z-0}). Using formulas (\ref{E}) and (\ref{p}) and substituting
\begin{equation}
\alpha_0=\frac{\sqrt{4-2\xi^2}}{2\xi},
\end{equation}
we obtain
\begin{equation}
\label{Edouble_xi}
     E(\phi_0)={}-\frac{\left(\xi ^2-2\right)}{3 \xi
   ^2} \left(3 C_1 B_2 \sqrt{4-2
   \xi ^2} \xi-3
   C_2 B_1 \sqrt{4-2 \xi ^2}
   \xi +8 B_1 B_2
   \left(2 \xi ^2-1\right)\right).
\end{equation}

The pressure is as follows
\begin{equation}
\label{Pdouble}
\begin{split}
P(\phi_0)=&-\frac{\left(\xi ^2-2\right)}{3 \xi
^2}\left(B_2e^{-\frac{t \sqrt{4-2 \xi ^2}}{\xi
   }}  \left(B_2\left(8
   \xi ^2-3 t \sqrt{4-2 \xi ^2} \xi
   -4\right) -3 C_2 \xi
   \sqrt{4-2 \xi ^2}\right)+\right.\\
+&\left.B_1
   e^{\frac{t \sqrt{4-2 \xi ^2}}{\xi }}
   \left(3 C_1 \sqrt{4-2 \xi ^2}
   \xi +B_1 \left(8 \xi ^2+3 t
   \sqrt{4-2 \xi ^2} \xi
   -4\right)\right)\right).
\end{split}
\end{equation}

\subsubsection{Energy Density and Pressure for real $\alpha$}
\label{pressure-real}
 As we have seen in Sect. 3.1.3 for some values of parameters $\xi$
and $c$ eq.~(\ref{5c}) has real roots. We denote as $p_m$ the values
of $p_\alpha$ for real $\alpha=m$,
\begin{equation}
 p_m=m^2\left(\xi^2-2+2\xi^2m^2\right),
\end{equation}
where $\xi^2$ is given by (\ref{equxi}).
For different values of $c$ the function $p_m$ is  presented in Fig.~\ref{p_m}.

\FIGURE{\centering
\includegraphics[width=40mm]{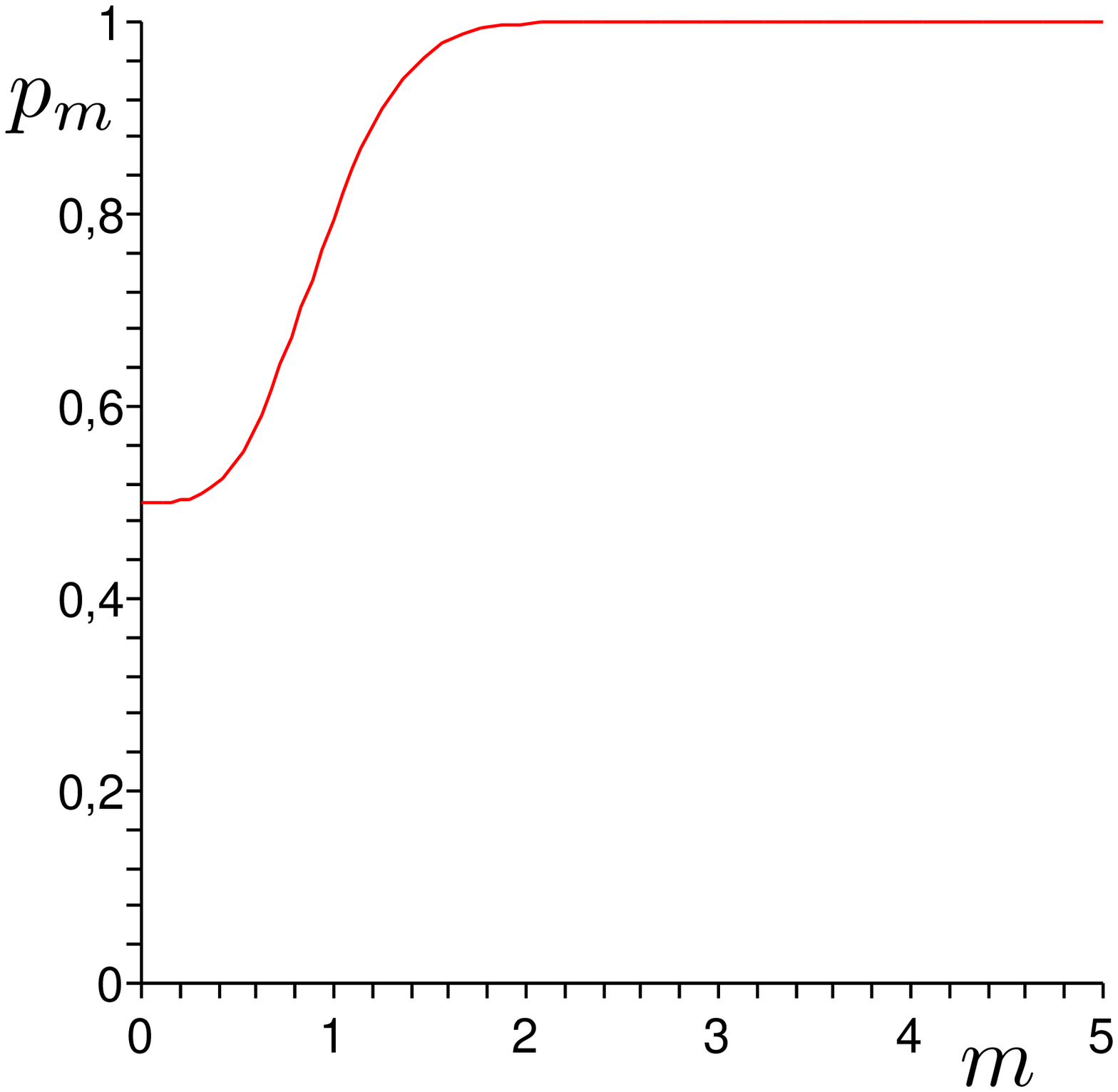} \ \ \ \ \
\includegraphics[width=40mm]{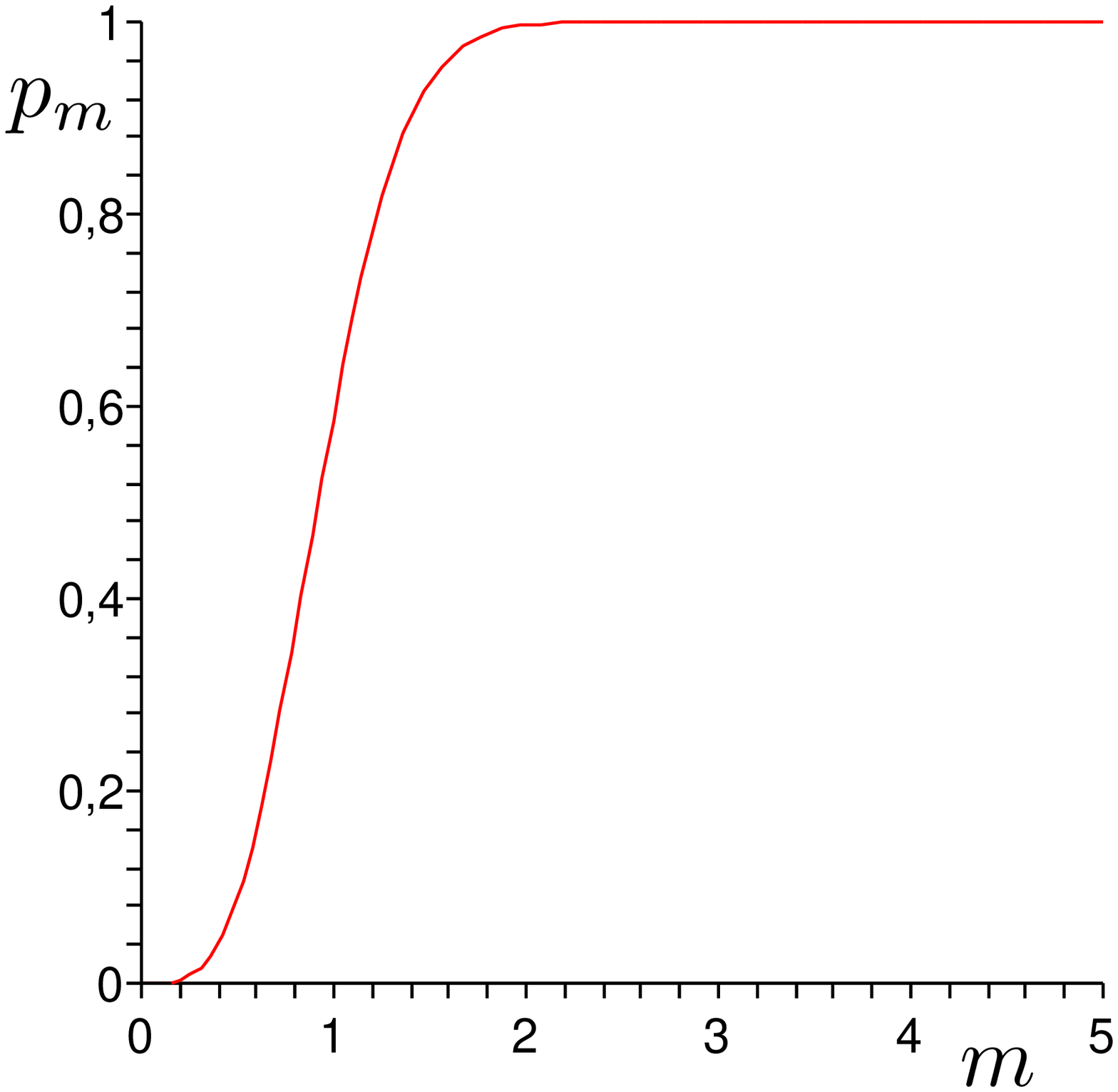} \ \ \ \ \
\includegraphics[width=40mm]{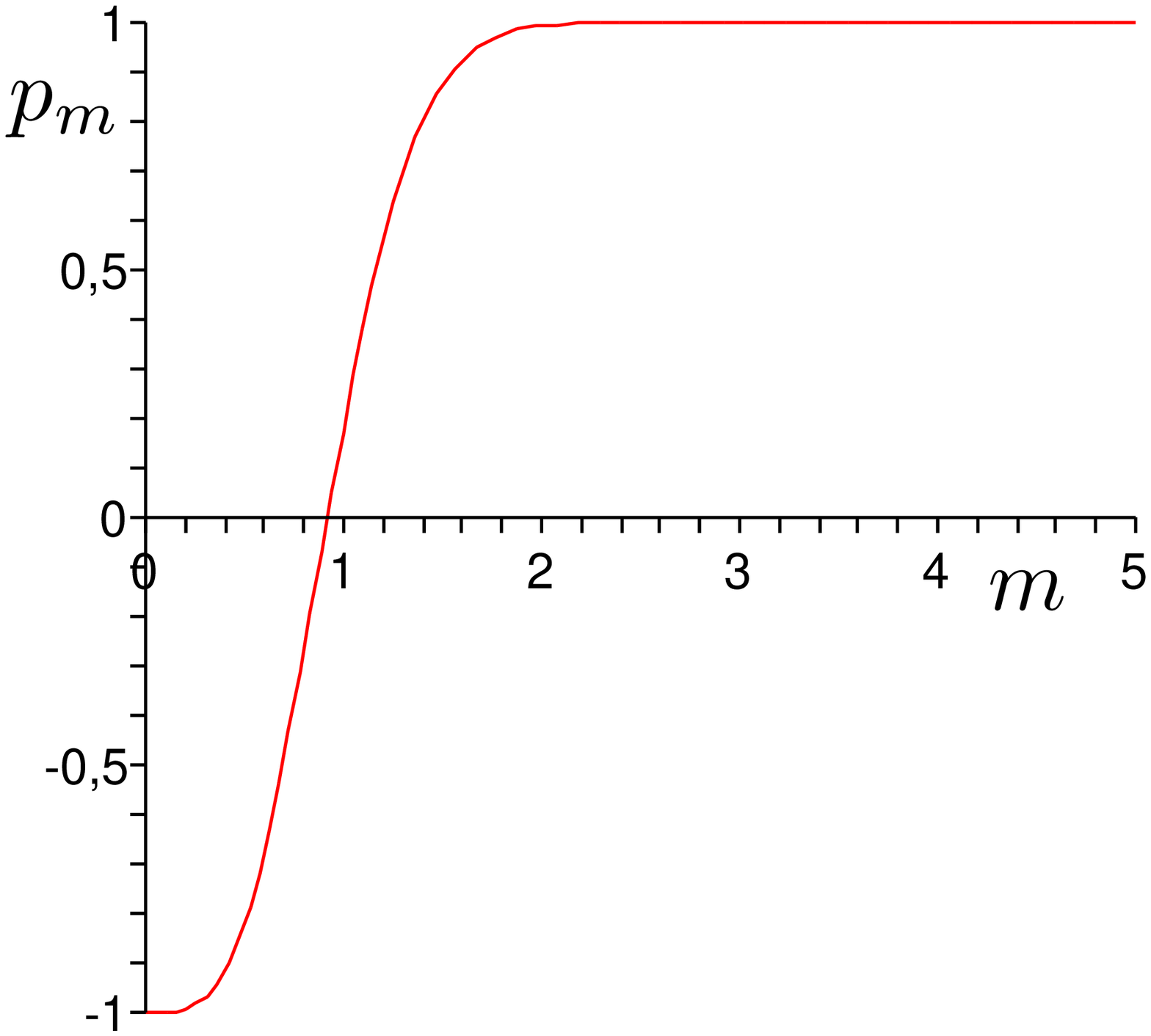}
\caption{ The dependence of $p_m$ on $m$ at $c=1/2$ (right), $c=1$
(center) and $c=2$ (left). } \label{p_m}}

If and only if $c>1$, then there exists the interval of
$0<m^2<m^2_{max}$ on which $p_m<0$. Some part of this interval is not
physical because on this part $g(m^2,c)<0$. The straightforward
calculations show that
\begin{equation}
 p_m=-\frac{\partial g(m^2,c)}{\partial m^2}.
\end{equation}
 Therefore, at the point
\begin{equation}
m_{max}^2=-\frac12-\frac12 W_{-1}\left(-\frac{e^{-1}}{c}\right),
\end{equation}
we obtain $p_m(m_{max})=0$. We conclude that for $c>1$ and
$\xi^2<\xi^2_{max}$ we have two positive roots of (\ref{5c}): $m_1$
and $m_2>m_1$, with $p_{m_1}<0$ and $p_{m_2}>0$.

The energy density and the pressure for solutions with real $\alpha$
one can calculate using formulas (\ref{Egen}) and (\ref{P-gen}) and
results are presented in
 Table 1.
\TABLE{\label{T-real}
  \centering
  \begin{tabular}{|c|c|c|}
\hline
  $\phi$ & $E$ & $P$ \\
  \hline
  $e^{\pm mt}$ &  0  & $p_me^{\pm 2m t}$\\ \hline
  $\sinh(mt)$ & $p_{m}/2$ & $p_{m}\cosh(2mt)/2$\\ \hline
  $\cosh(mt)$ & ${}-p_{m}/2$ & $p_{m}\cosh(2mt)/2$\\
\hline
\end{tabular}
  \caption{Solutions, densities of energy  and
 pressures for real $\alpha$}}

We see from Table 1 that odd solutions are physically meaningful,
$E>0$, if $p_m$ is positive.  Fig.\ref{p_m} shows that odd solutions
are physical for $c< 1$ and any $\xi^2$ and for $c>1$ only for
$m^2>m_{max}^2$. The  pressure corresponding to this solution is always
positive.

Even solutions are physically meaningful
 if $p_m$ is
negative.   Therefore, even solutions are physical only for $c>1$
and $m^2<m_{max}^2$. The  pressure corresponding to this solution
is always negative. The equation of the state parameter for this
solution is \begin{equation}w=-\cosh (2mt)<-1. \end{equation}

\subsubsection{Energy Density and Pressure for pure imaginary $\alpha$}
As we have seen in Sect.3.1.4 for some values of parameters $\xi$
and $c$ eq.~(\ref{5c})  has only pair of pure imaginary
 roots.
These solutions correspond to
\begin{equation}
\label{sin-sol}
   \phi=C\sin(\mu(t-t_0)).
\end{equation}

On the  solutions (\ref{sin-sol}) the energy and pressure are given
by
\begin{eqnarray}
\label{E-s}
E(C\sin(\mu(t-t_0)))&=&\frac{C^2}{2}p_\alpha\equiv\frac{C^2}{2}\pi_{\mu},\\
\label{p-sin} P(C\sin(\mu(t-t_0)))&=& -\frac{C^2}{2}\pi_\mu\cos(2\mu
(t-t_0)),
\end{eqnarray}
where
\begin{equation}
\label{E-mu} \pi_\mu\equiv p_{i\mu}=\mu^2\left(2+2\xi^2\mu^2-\xi^2\right),
\end{equation}
where $\xi^2$ is given by (\ref{5da}). $\pi_{\mu}$ as a function
of $\mu$ for different values of $c$ is presented in
Fig~(\ref{pi_mu}).

\FIGURE{ \centering
\includegraphics[width=45mm]{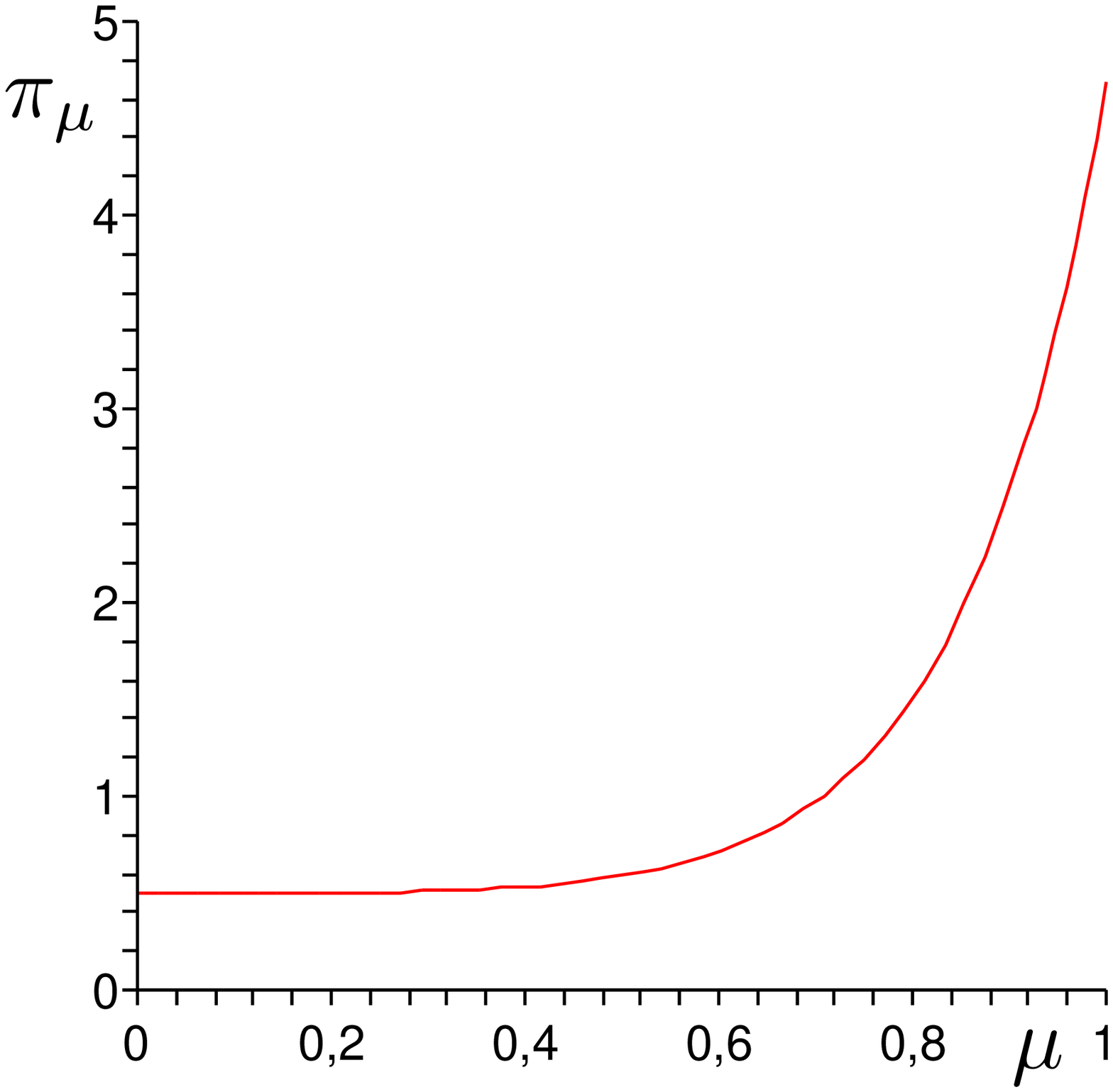} \ \ \ \ \
\includegraphics[width=45mm]{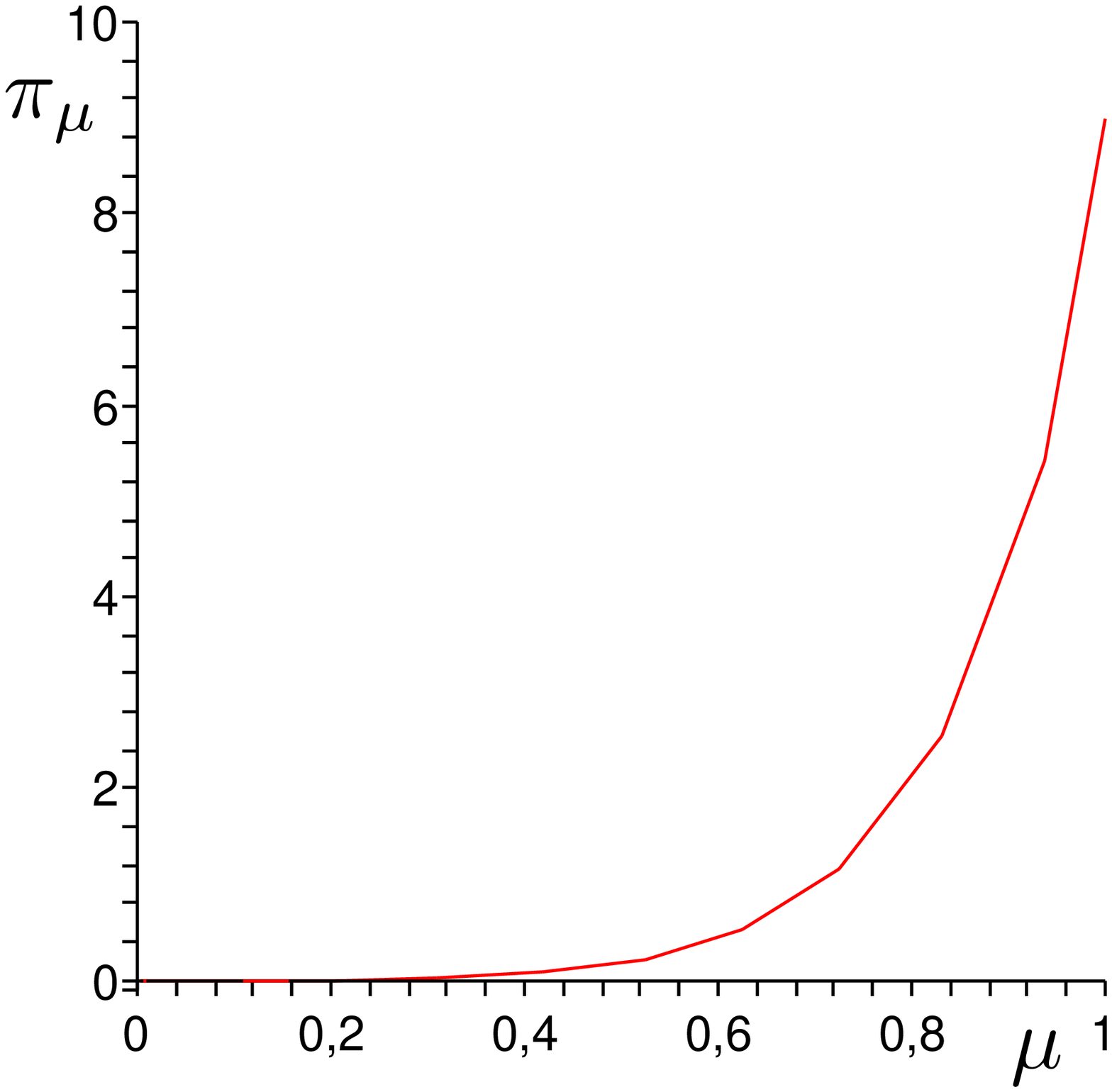} \ \ \ \ \
\includegraphics[width=45mm]{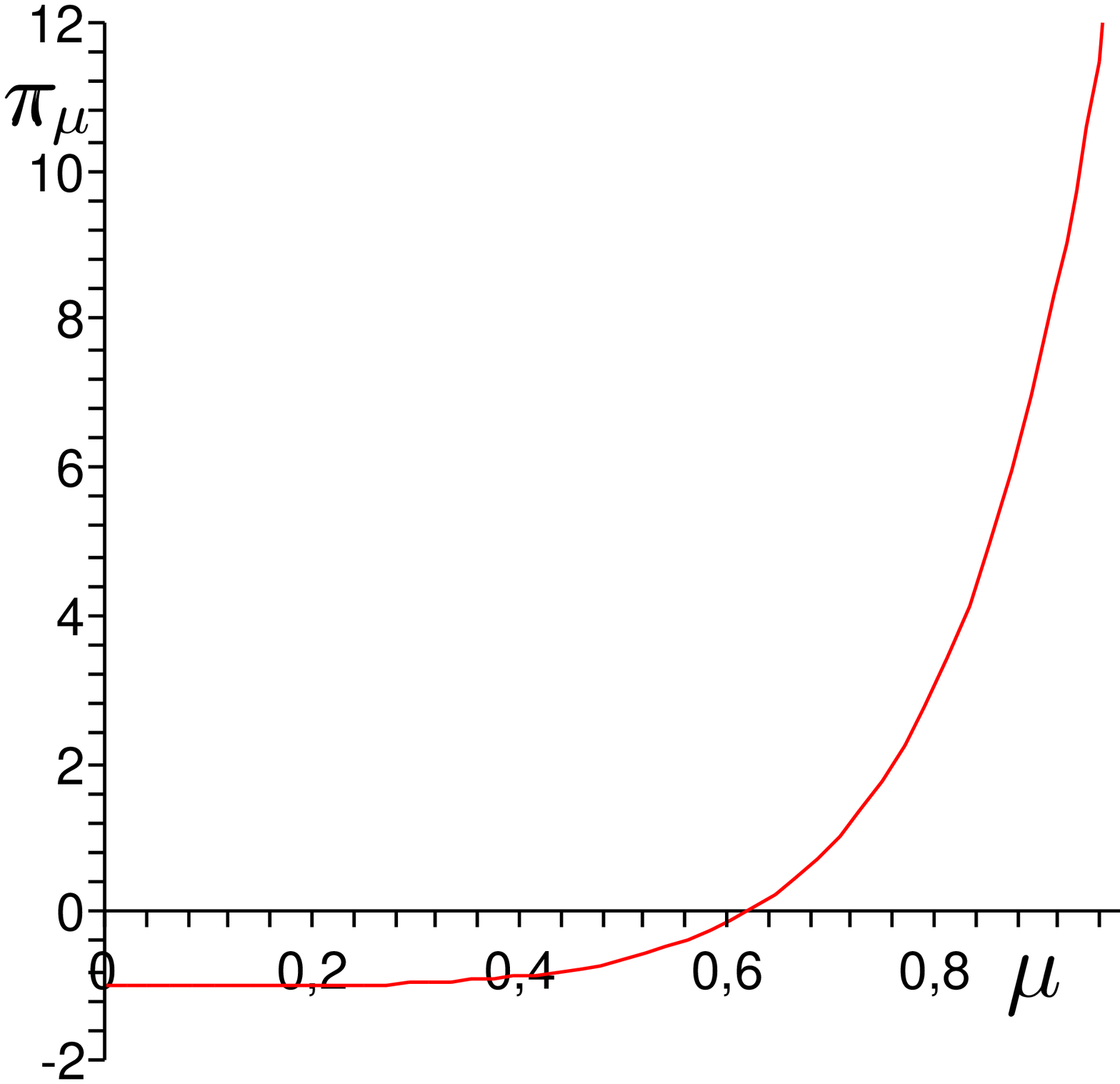}
\caption{ The dependence of $\pi_{\mu}$ on $\mu$ at $c=1/2$
(right), $c=1$ (center) and $c=2$ (left). } \label{pi_mu}}

Note that $\pi_\mu$ for $c\leqslant 1$ is positive, however  for
$c>1$ is positive only for $\mu^2\geqslant\mu_{min}^2=\frac12
-\frac{1}{\xi^2}$. The equation of the state parameter for this
solution is \begin{equation}w=-\cos (2\mu(t-t_0)),
~~\Rightarrow~~~|w|\leqslant 1. \end{equation}

\subsubsection{Energy and Pressure in the case $c=1$}

The energy density and pressure for solutions (\ref{equsol}):
\begin{equation}
\label{linearphi} \phi_1(t)=C_1t+C_0,
\end{equation}
where $C_0$ and $C_1$ are arbitrary constants, are as follows
\begin{equation}
    E(\phi_1)=\left(\frac{\xi^2}{2}-1\right)C_1^2,\qquad
    P(\phi_1)=\left(\frac{\xi^2}{2}-1\right)C_1^2
\end{equation}
and the state parameter $w\equiv P/E=1$.

The straightforward calculations show that pressure and energy
density for more general solutions
\begin{equation}
\label{phiexpsumlin}
    \phi=\sum_{n=1}^{N} \tilde{C}_n e^{\alpha_n t}+\phi_1
\end{equation}
are
\begin{equation}
    E\left(\sum_{n=1}^{N} \tilde{C}_n e^{\alpha_n t}+\phi_1\right)
    =E\left(\sum_{n=1}^{N} \tilde{C}_n e^{\alpha_n
    t}\right)+E\left(\phi_1\right)
\end{equation}
and
\begin{equation}
    P\left(\sum_{n=1}^{N} \tilde{C}_n e^{\alpha_n t}+\phi_1\right)
    =P\left(\sum_{n=1}^{N} \tilde{C}_n e^{\alpha_n
    t}\right)+P\left(\phi_1\right).
\end{equation}

Let, for example,
\begin{equation*}
  \phi(t)=\cosh(mt)+C_1t+C_0.
\end{equation*}
The corresponding energy density and pressure are:
\begin{equation*}
  E=-\frac{1}{2}p_m+\left(\frac{\xi^2}{2}-1\right)C_1^2, \qquad
  P=\frac{1}{2}p_m\cosh(2mt)+\left(\frac{\xi^2}{2}-1\right)C_1^2.
\end{equation*}
In the case $\xi^2=2$ the root $\alpha=0$ is a double root of
eq.~(\ref{5c}), so eq.~(\ref{1c}) has solutions~(\ref{equsol4}):
\begin{equation*}
    \phi_2=C_3t^3+C_2t^2+C_1t+C_0.
\end{equation*}

We obtain:
\begin{equation*}
 E(\phi_2)=4\left(3C_1C_3-C_2^2-6C_3^2\right), \qquad
P(\phi_2)=72C_3^2t^2+48C_2C_3t+4\left(C_2^2+3C_1C_3-6C_3^2\right).
\end{equation*}

\subsubsection{Energy and pressure for complex
$\alpha=r+i\nu$}

For a decreasing solution \begin{equation}\label{dec-c} \phi(t)
=e^{-rt}\cos(\nu t ) \end{equation}we have
\begin{equation}\label{dec-E} E\left(e^{-rt}\cos(\nu t)\right)=0,
\end{equation}\begin{equation}\label{dec-P} P\left(e^{-rt}\cos(\nu
t)\right)=\frac{e^{-2rt }}{4}\left(p_{\alpha}e^{2i\nu
t}+p_{\alpha^*}e^{-2i\nu t}\right). \end{equation}

 Using general formulas
(\ref{Egen}) and (\ref{P-gen}) it is easy to find the energy and
pressure for even and odd real solutions with complex $\alpha$. For
example the energy and pressure for the even solution $\phi=\cosh(r
t)\cos(\nu t)$ are as follows:
\begin{equation}
\label{Ecoshcos}
 E(\cosh(rt)\cos(\nu t))=-\frac{1}{8}
 \left(p_{\alpha}+p_{\alpha^*}\right),
\end{equation}
\begin{equation}
\label{Pcoshcos}
 P(\cosh(rt)\cos(\nu t))=\frac{p_\alpha}{16}
 \left(e^{-2\alpha
t}+e^{2\alpha t}\right)+\frac{p_{\alpha^*}}{16}\left(e^{-2\alpha^*
t}+e^{2\alpha ^*t}\right).
 \end{equation}
The equation of state parameter on the even solution is
\begin{equation}w=-\frac{p_\alpha +p_\alpha^*}{p_\alpha \cosh
(\alpha t) +p_\alpha^*\cosh (\alpha^* t)}. \end{equation}
\subsection{Local Field Representation}
\subsubsection{Weierstrass product and  Mode Decomposition for the
Action}

As in \cite{AV} we can present $F(-\square)$ in the action
(\ref{ACTION}) as the Ostrogradski Representation. To this purpose
let us construct the Weierstrass product for the function $F(z)$
of a complex variable $z$. Let us recall that a complex function
$R(z)$ such that its logarithmic derivative $R'(z)/R(z)$ is a
meromorphic function regular in the point $z=0$, has  simple poles
and satisfies $|R'(z)/R(z)|<C$, $z\in \Gamma_n$, $n=1,2,...$, can
be presented as
\begin{equation}
\label{WP} R(z)=R(0)e^{\frac{R'(0)}{R(0)}z}
\prod\left(1-\frac{z}{z_k}\right)e^{z/z_{k}}.
\end{equation}
$\Gamma_n$,  $n=1,2,...$ is a set of special closed contours
$\Gamma_n$ such that the point $z=0$ is in all $\Gamma_n$,
$\Gamma_n$ is in $\Gamma_{n+1}$, and $S_n/d_n\leq C$, where $S_n$
is a length of the contour $\Gamma_n$, and $d_n$ is its distance
from zero \cite{Shabat}.

In the case of a more week requirement $|R'(z)/R(z)|<M|z|^p$ $z\in
\Gamma_n$, $n=1,2,...$ we have \begin{equation}\label{WP-d}
R(z)=e^{f(z)} \prod\left(1-\frac{z}{z_k}\right)e^{Q_{k}(z)},
\qquad Q_{k}(z)=\sum_{l=1}^{p}
\frac1l\left(\frac{z}{z_k}\right)^l, \end{equation}where $f(z)$ is
an entire function.

In the case of $R=F$ given by (\ref{F}) the Weierstrass product
can be written in the form
\begin{equation}
F(\alpha^2)=e^{f(\alpha^2)}\prod_n (\alpha^2-\alpha_n^2).
\label{5cW}
\end{equation}
The function $f(z)$ in our case is
\begin{equation}
f(z)=A+\beta z,
\end{equation}
where constants $A$ and $\beta$ are determined by $\xi$ and $c$.
It will be shown that the equations of motion do not depend on
values of $A$ and $\beta$.

It is convenient to pick out real roots in (\ref{5cW}) and combine
the complex conjugated roots:
\begin{equation}
F(\alpha^2)=e^{A+\beta \alpha^2}\prod (\alpha^2-m_k^2)\prod
(\alpha^2-\alpha_n^2)(\alpha^2-\alpha_n^{*2}), \label{5cW1}
\end{equation}
where $m_k$ denote real roots. In Subsection 3.1 we have found the
cases when real roots do exist.

In the case of simple roots the Lagrangian up to a total
derivative can be presented as a sum of an infinite number of
fields \cite{Ost}--\cite{Volovich}
\begin{equation}
{\cal L}=\frac12 \phi F(\partial ^2 )\phi \sim \frac12 \sum
\left[\epsilon _n \psi_n e^{f(\partial^2)}(-\partial
^2+\alpha_n^2)\psi_n +c.c.\right], \label{l-inf-pairs}
\end{equation}
where $\sim $ means equivalence up to a total derivative,
$\epsilon_n$ are constants. It is the Ostrogradski representation.
Note that for complex roots $\psi_n$ are complex.

\subsubsection{Mode Decomposition for Energy Density and Pressure}

It is instructive to present the formula for energy and pressure
obtained in  section 3.2. in terms of $\psi$ fields. All
considerations below take place for nondegenerate roots.

According to a general procedure of construction of the energy and
pressure  we write a generalization of (\ref{l-inf-pairs})
to a non-flat case

\begin{equation}
{\cal L}_g= \sum {\cal L}_g(\psi _n),\qquad {\cal L}_g(\psi _n)=
 \frac{\epsilon _n }{2}\sqrt{-g}\psi_n
e^{f(-\square_g)}\left(\square_g+\alpha_n^2\right)\psi_n,\label{lg-inf-fields}
\end{equation}
and find \bea \label{Et-psi} E_\psi=\sum_n E_n,\qquad
E_n=\frac{\epsilon _n}{2}\left(\dot{\psi_n}^2-\alpha_n^2
\psi_n^2\right)e^{f( \alpha_n^2)}\label{E-psin},\\
\label{Pt-psi} P_\psi=\sum_n P_n,\qquad P_n= \frac{\epsilon
_n}{2}\left(\dot{\psi_n}^2+\alpha_n^2
\psi_n^2\right)e^{f(\alpha_n^2)}.\label{P-psin} \eea

The E.O.M. for $\psi_n$ is
\begin{equation}
(\partial^2-\alpha_n^2)\psi_n=0
\end{equation}
and its solutions are
\begin{equation}
\psi_n=A_ne^{\alpha_nt}+B_ne^{-\alpha_nt}.
\label{psi-n}
\end{equation}
For solutions (\ref{psi-n}) we obtain
 \bea
\label{E-exp-psin} E_\psi&=&2\sum_n
\alpha_n^2\epsilon _nA_nB_n e^{\beta \alpha_n^2},\\
P_\psi&=&\sum _n
\epsilon_n\alpha_n^2\left(A^2_ne^{2\alpha_nt}+B_n^2e^{-2\alpha_nt}\right)e^{\beta
\alpha_n^2}. \label{P-exp-psin} \eea

On the other hand according to (\ref{Egen}) and (\ref{P-gen}) we
have \bea \label{E-old}
E&=&-2\sum_nA_nB_n\alpha_n^2\left(\xi^2-2+2\xi^2\alpha_n^2\right),\\
\label{P-old} P&=&\sum_n\left(A_n^2e^{-2\alpha_n
t}+B_n^2e^{2\alpha_n
t}\right)\alpha_n^2\left(\xi^2-2+2\xi^2\alpha_n^2\right). \eea
Comparing (\ref{E-exp-psin}),(\ref{P-exp-psin}) and
(\ref{E-old}),(\ref{P-old}) and using equation (\ref{5c}) we
obtain that
\begin{equation}
    E=E_\psi, \qquad P=P_\psi
\end{equation}
if and only if
\begin{equation}
\label{eq-epsilon} \epsilon_n=-(2ce^{-2\alpha_n^2}+\xi^2)e^{-\beta
\alpha_n^2},
\end{equation}
that is in accordance with general formula for $\epsilon_n$
\cite{PaisU,AV}. Note that we consider only simple roots
$\alpha_n$.

\subsection{Finite Order Derivative Approximation}

\subsubsection{Two types of approximations}
There are two different types of finite order derivative
approximations:
\begin{itemize}
\item  a direct finite order derivative approximation
\begin{equation}
L_{2k}(\alpha^2)\equiv \sum_{n}^k a_n \alpha^{2n}; \label{apr-f}
\end{equation}

\item an approximation by a finite number of terms in the
Weierstrass product
\begin{equation}
L^{(W)}_{2k}(\alpha^2)\equiv f(\alpha^2)\prod_{n}^k (\alpha^2-
\alpha^2_n). \label{apr-W}
\end{equation}
We label roots so that $|\alpha _n|\leqslant|\alpha_{n+1}|$.
\end{itemize}
Locations of  roots of the characteristic equation in the complex
$\alpha$-plane are presented in Figure~\ref{figure_roots}. One can
see that structures of roots location for different values of
$\xi$ look similar.

\FIGURE{ \centering
\includegraphics[width=52.72mm]{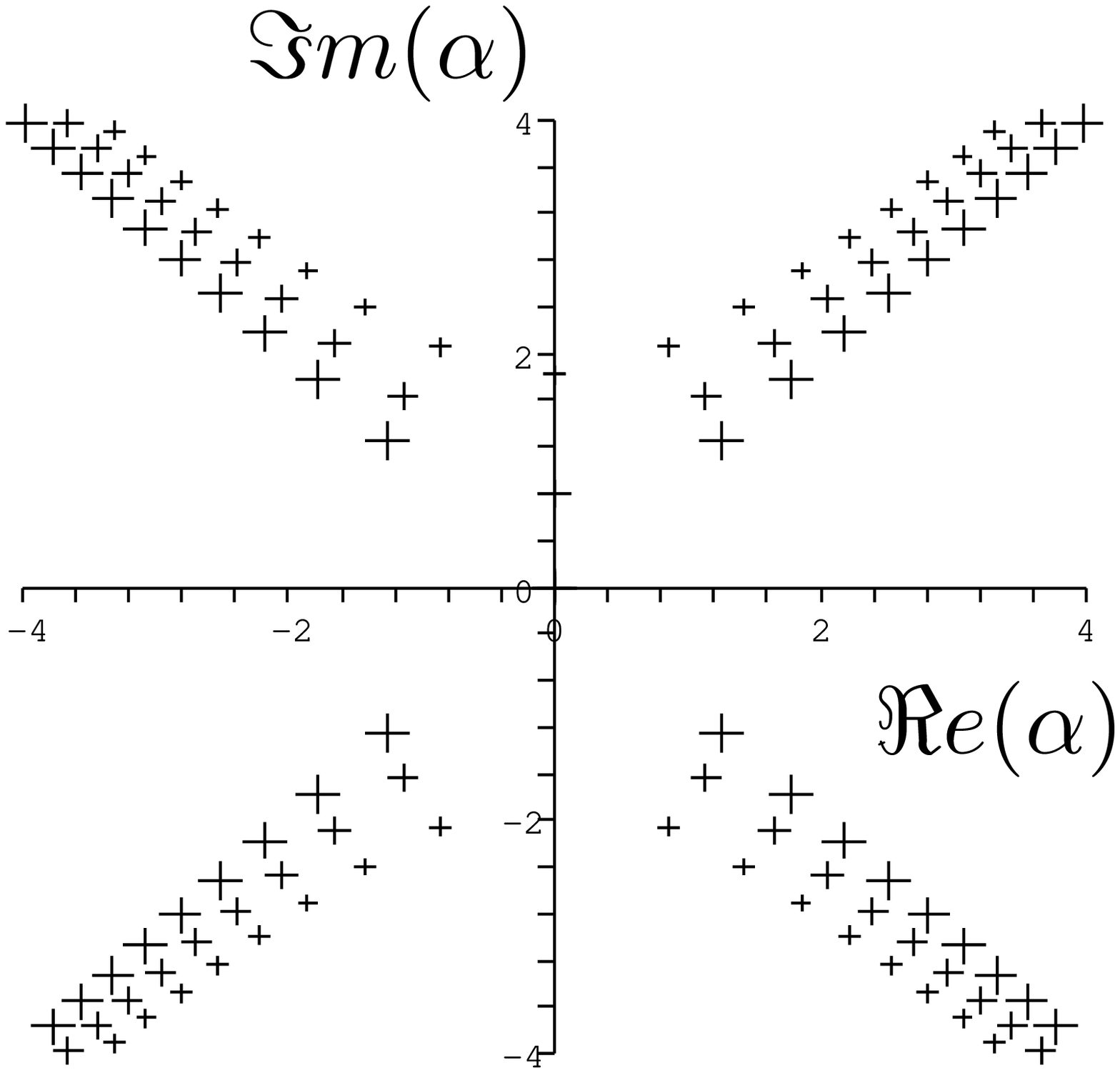}
\caption{Roots for $c=1$  and $\xi=0$ (big crosses), $\xi=2$
(middle crosses)  and $\xi=15$ (small crosses).}
\label{figure_roots}} Let us consider the most simple case
$\xi=0$, $c=1$, that corresponds to
\begin{equation}
L(\alpha^2)\equiv L_{0,1}(\alpha^2)= 1- e^{-2\alpha ^2}. \label{0,1}
\end{equation}

In this case all roots can be written explicitly
\begin{equation}
\label{Non2} 1-e^{-2\alpha ^2}=2\alpha ^2e^{-\alpha
^2}\prod_{j=1}^{\infty} \left(1+\frac{\alpha ^4}{\pi^2
j^2}\right).
\end{equation}
Note that all conclusions can be generalized on the case of
arbitrary $\xi$ and $c$ such that all roots $\alpha_n$ are simple.

\subsubsection{Direct Finite Order Approximation}

In the second order derivative approximation we sho\-uld keep in
(\ref{1c}) with $c=1$, $\xi=0$ only the second derivatives
\begin{equation}
2\partial ^2  \phi =0. \label{3-2}
\end{equation}
This equation has the following solutions
\begin{equation}
 \phi=C_1t+C_2 \label{3-2-s}.
\end{equation}

These solutions correspond to roots $\alpha_1=0$.

In the fourth order derivative approximation eq.~(\ref{5c}) at $c=1$
and $\xi=0$ is as follows
\begin{equation}
\label{xi4}
    2\alpha^2-2\alpha^4=0
\end{equation}
and has two solutions:
\begin{equation}
\alpha^2=0, \qquad \alpha^2=1.
\end{equation}

We see that in the fourth order direct approximation the
approximate equation (\ref{xi4}) has a root ($\alpha=1$) that is
absent in the initial equation (\ref{5c}).

The similar situation takes place for higher order approximations.
The characteristic equation of the direct n-order approximation
contains several artificial roots. The appearance of these roots
is related to artificial roots of polynomial approximations of the
function $f$ in the Weierstrass product (\ref{5cW}). In
Figure~\ref{roots} we plot all roots of an n-order polynomial
approximation of function $L_{0,1}(\alpha^2)$ for n=20 and 40. We
see that these polynomials have true roots as well as artificial
roots that go to infinity when the order of the approximate
polynomial increases.

\FIGURE{ \centering
\includegraphics[width=43mm]{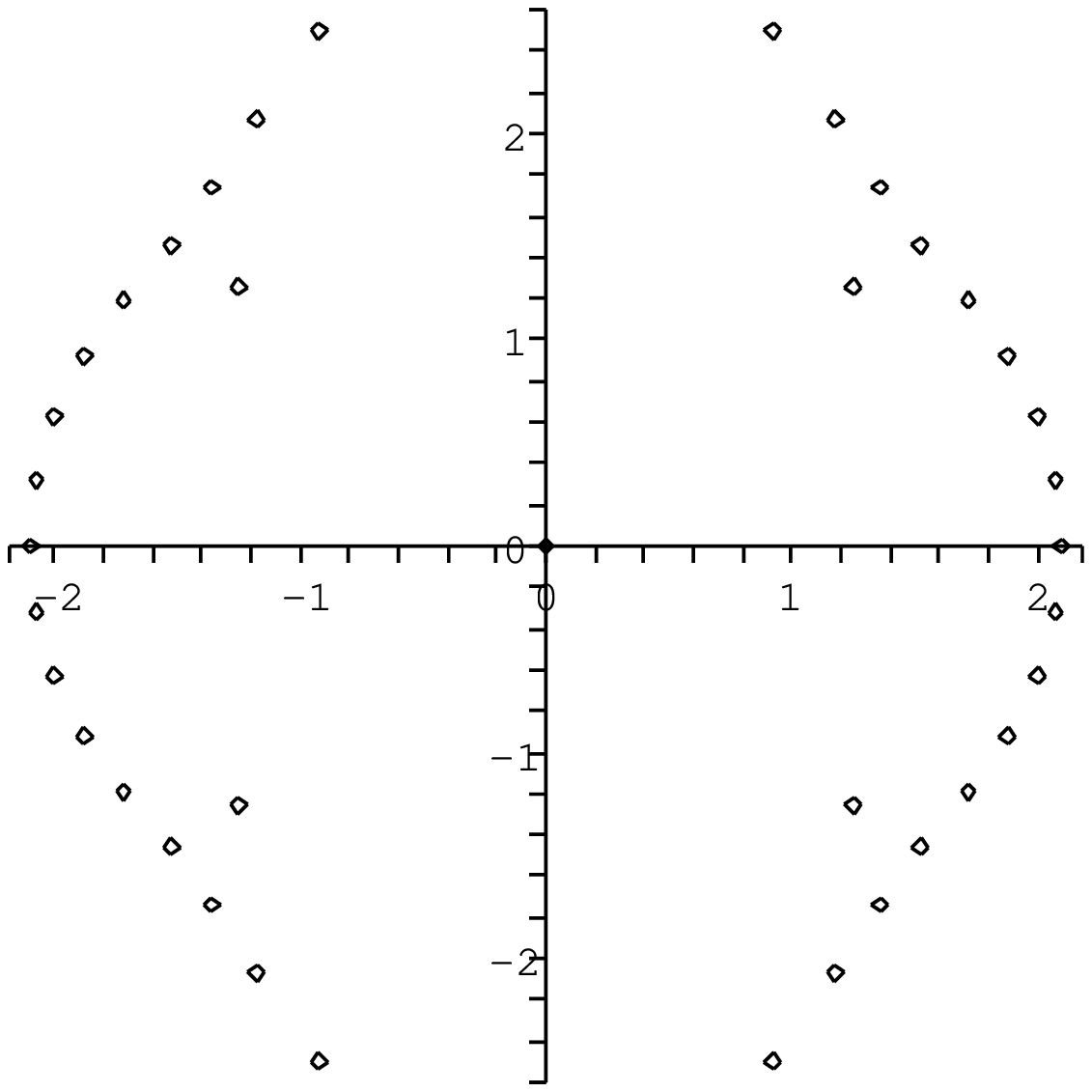} A.\
\includegraphics[width=43mm]{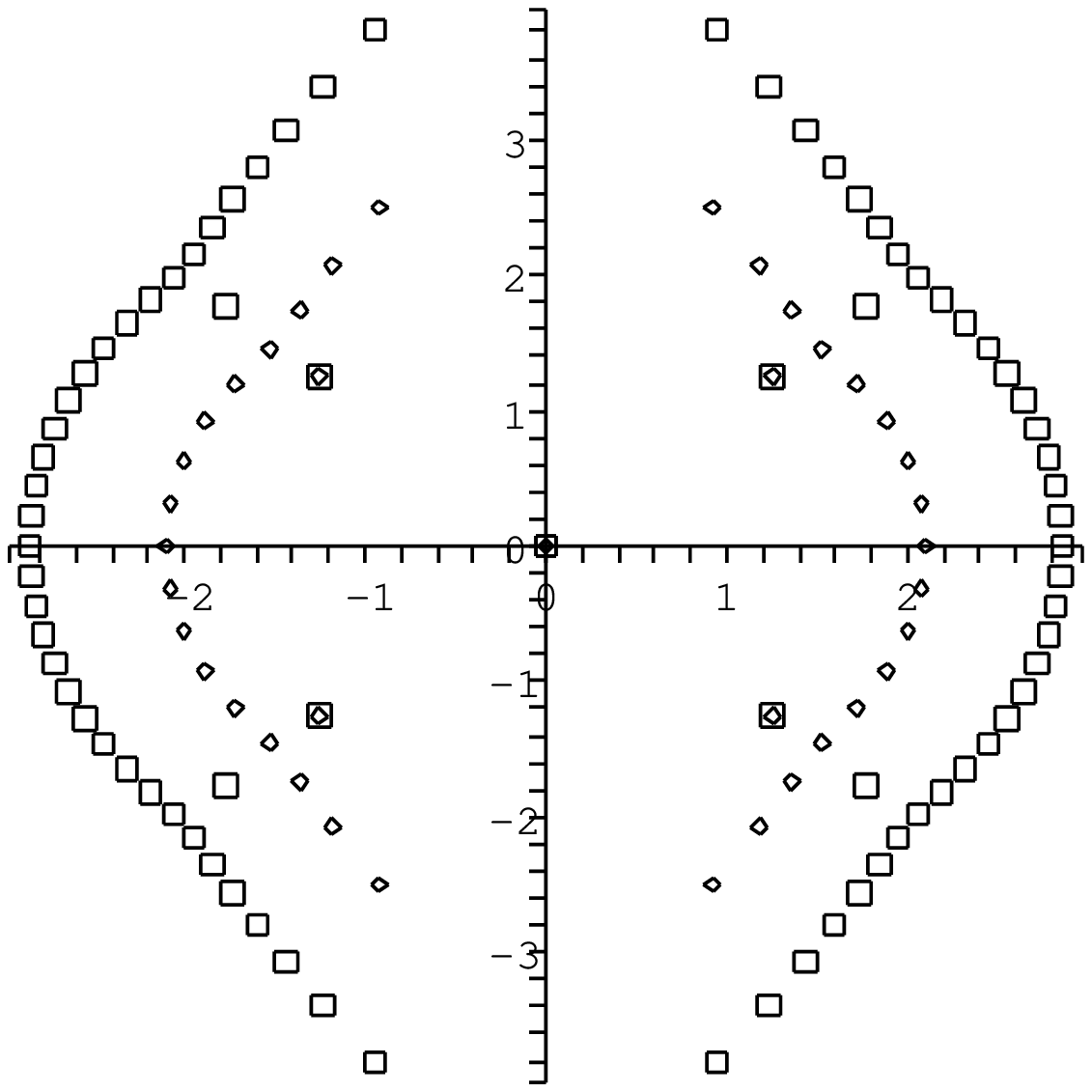} \ B.\
\includegraphics[width=43mm]{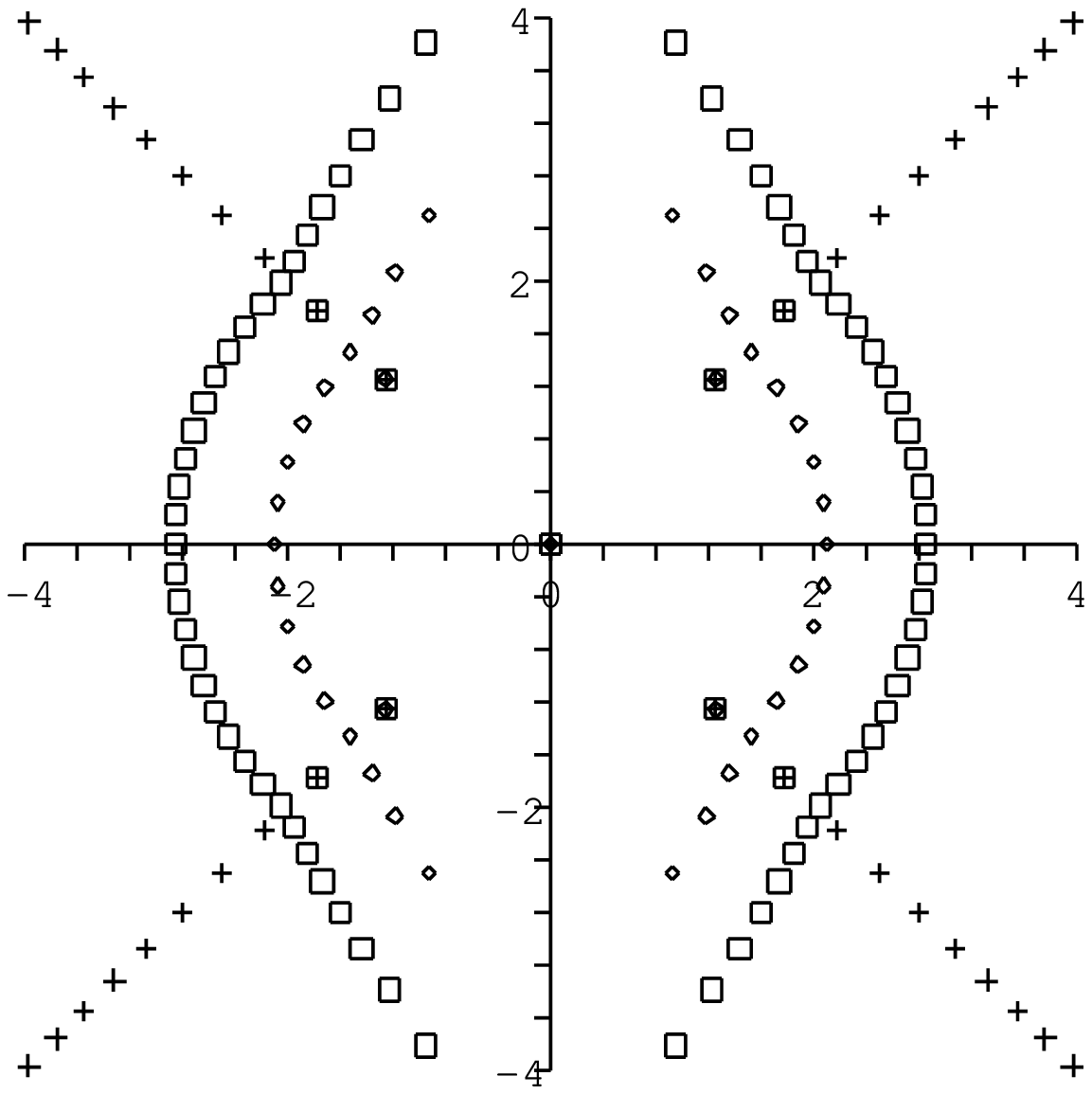}~~C.
\caption{A. Roots of the 20-th order approximation
$L_{0,1;10}(\alpha)$ of the function $L_{0,1}(\alpha^2)$ depicted by
small diamonds. There are  one root in the origin, 4 roots located
on a "cross" and  15 roots on  two bow-like curves. B. Roots of the
40-th order approximations, $L_{0,1;20}(\alpha)$, and roots of
$L_{0,1;10}(\alpha)$
 depicted by small boxes and small diamonds, respectively.
 $L_{0,1,20}(\alpha)$ has one root in the origin, 8 roots located on a
"cross" and 4 of them coincide with true roots  of
$L_{0,1;10}(\alpha)$. Artificial roots of  $L_{0,1,20}(\alpha^2)$
are located  on  bigger "bows" as compare with artificial roots of
$L_{0,1,10}(\alpha^2)$. C. Exact roots of $L_{0,1}(\alpha)$ are
located on a cross and depicted by small crosses. The roots of the
$L_{0,1;10}(\alpha)$ and $L_{0,1;20}(\alpha)$, are depicted by
small diamonds  and  small boxes, respectively, as in A and B.}
\label{roots}}

\subsubsection{Finite Order Weierstrass Product Approximation}
 We  consider an
approximation to $L_{0,1}(\alpha^2)$ that keeps only a finite number
of terms in the Weierstrass product
\begin{equation}
L^{(W)}_{0,1,2+4k}(\alpha^2)\equiv 2\alpha^2e
^{-\alpha^2}\prod_{j=1}^k \left(1+\frac{ \alpha^4}{\pi^2j^2}\right).
\label{apr-W-L}
\end{equation}
We call this approximation as (2k+1)-roots approximation. The
corresponding Lagrangian is
\begin{equation}
{\cal L}^{(W)}_{0,1,2+4k}\equiv \frac12\phi \partial ^2e
^{-\partial^2}\prod_{j=1}^k \left(1+\frac{
\partial^4}{\pi^2j^2}\right)\phi. \label{apr-W-Lagr}
\end{equation}

Let us consider one root  approximation
\begin{equation}
{\cal L}^{(W)}_{0,1,2}\equiv \frac12 \phi \partial ^2e
^{-\partial^2}\phi. \label{apr2}
\end{equation}
E.O.M. is
\begin{equation}
\partial ^2e ^{-\partial^2}\phi=0
\end{equation}
and it has an unique solution (the same as E.O.M. of 2-nd order
derivative approximation)
\begin{equation}
\phi=At+C.
\end{equation}

The next approximation is 3 roots approximation
\begin{equation}
{\cal L}^{(W)}_{0,1,6}\equiv \frac12\phi \partial ^2e ^{-\partial^2}
\left(1+\frac{ \partial^4}{\pi^2}\right)\phi. \label{apr-W-6}
\end{equation}

E.O.M. is
\begin{equation}
\partial ^2\left(1+\frac{ \partial^4}{\pi^2}\right)e ^{-\partial^2}\phi=0
\qquad \Leftrightarrow \qquad
\partial ^2\left(1+\frac{ \partial^4}{\pi^2}\right)\phi=0
\end{equation}
and it has the following solutions
\begin{equation}
\phi_6(t)=A_1t+\delta_1+A_2e^{\sqrt{\frac{\pi}{2}}\,t}\sin
(\sqrt{\frac{\pi}{2}}\,t+\delta_2)
+A_3e^{-\sqrt{\frac{\pi}{2}}\,t}\sin
(\sqrt{\frac{\pi}{2}}\,t+\delta_3),
\end{equation} where $A_k$ and
$\delta_k$ are arbitrary constants.

The next approximation contains 4 extra roots
\begin{equation}
{\cal L}^{(W)}_{0,1,10}\equiv \frac12\phi \partial ^2e
^{-\partial^2} \left(1+\frac{ \partial^4}{\pi^2}\right)
\left(1+\frac{ \partial^4}{4\pi^2}\right)\phi. \label{apr-W-8}
\end{equation}
E.O.M. is
\begin{equation}\partial ^2\left(1+\frac{ \partial^4}{\pi^2}\right)\left(1+\frac{
\partial^4}{4\pi^2}\right)
e ^{-\partial^2}\phi=0 \end{equation}and it has the following
solutions
\begin{equation}\phi_{10}(t)=\phi_{6}(t)+A_4e^{\sqrt{\pi}\,t}\sin
(\sqrt{\pi}\,t+\delta_4) +A_5e^{-\sqrt{\pi}\,t}\sin
(\sqrt{\pi}\,t+\delta_5). \end{equation} It is obvious that
solutions of the Weierstrass approximate equations reproduce a
finite number of modes of the full equation.

If we restrict ourself to decreasing solutions
\begin{equation}\phi_{10,\mbox {decr}}(t)=\delta_1+A_3e^{-\sqrt{\frac{\pi}{2}}\,t}
\sin \left(\sqrt{\frac{\pi}{2}}\,t+\delta_3\right)
+A_5e^{-\sqrt{\pi}\,t}\sin (\sqrt{\pi}\,t+\delta_5)
\end{equation}
we see that the last term can be ignored as compared with term first two terms.

We can see that the Weierstrass product approximation is more
preferable than direct approximation, because there is no problem
with extra roots. In the next section we will use the Weierstrass
product approximation to construct local cosmological models.

\section{Non-flat Dynamics}
\subsection{Modified  Action}

The goal of this section is to consider the nonlocal model
(\ref{ACTION}) in the  Friedmann Universe. To consider the
dynamics in such a system we need to solve nonlinear Friedmann
equations (\ref{eomprho}), which represent hopelessly complicated
problem. From (\ref{eomprho}) we obtain the following nonlinear
integral equation in $H(t)$:
\begin{equation}
\dot
H={}-\frac{1}{2m_p^2}({\cal{P}}+{\cal{E}})=-\frac{1}{m_p^2}\left(
\frac{\xi^2}{2}(\partial\phi)^2-c\int_{0}^{1}\left(\partial
e^{(1+\rho){\cal D}} \phi\right) \left(\partial e^{(1-\rho) {\cal
D}}\phi\right) d \rho\right), \label{dH}
\end{equation}
where ${\cal D}\equiv {}- \partial_t^2- 3H(t)\partial_t$. There is
no method to solve eq.~(\ref{dH}), even if the function $\phi(t)$
is given.

As a first step to consider such a problem we would like to
construct some effective  exactly solvable models that can be
consider as an approximation to the nonlocal model.
  To do this we use the finite order
approximations, constructed by the Ostrogradski method. In other
words we choose a special solution of eq.~(\ref{1c}) and find the
corresponding Ostrogradski approximation in the flat space-time.
After we deform the obtained approximate model to the case of the
Friedmann Universe, assuming that exact solutions in the Friedmann
metric are coincide with exact solutions in the flat space-time.
Note that the similar assumption has been used in the papers
\cite{AKV,AJ,AKVtwofields} to construct effective local models
with exact solutions.

Our starting point is the Lagrangian (\ref{l-inf-pairs}). The
corresponding action in the non-flat space-time is as follows:
\begin{equation}
\label{action-in-background-mod} S_{Ostr.}=\int d^4x
\sqrt{-g}\left(\frac{m_p^2}{2}R+ \sum_n \left[\frac{\epsilon _n}{2}
\psi_n e^{\beta\alpha_n^2}(\square_g+\alpha_n^2)\psi_n
+c.c.\right]-{\cal V}(\psi_1,\dots,\psi_n)\right).
\end{equation}

If the fields $\phi_n$ depend only on time and the metric is a
spatially flat Friedmann metric, then we have the following
equation for $\psi_n$
\begin{equation}
\epsilon _n\left({}{\cal D}  +\alpha_n^2\right)\psi_n
-e^{-\beta\alpha _n^2}{\cal V}_{\psi_n}^\prime=0 \quad
\Leftrightarrow \quad (2ce^{-2\alpha_n^2}+\xi^2)\left({}{\cal D}
+\alpha_n^2\right)\psi_n+{\cal V}_{\psi_n}^\prime=0,
\label{eom-phi-H-mod}
\end{equation}
where ${\cal V}^\prime_{\psi_n}$ is a derivative of ${\cal V}$ on
${\psi_n}$. Note that form of ${\cal V}(\psi_1,\dots,\psi_n)$
depends on choose of special solutions $\psi_1$, $\dots$, $\psi_n$.
The form of ${\cal V}(\psi_1,\dots,\psi_n)$ is given below
(Subsection 4.4).

The energy and the pressure density in the Friedmann metric have the
form \bea \label{E-phi-H-mod} {\cal E}_{mod}&=&E_\psi+{\cal V},
\\
\label{PFr-mod} {\cal P}_{mod}&=&P_\psi-{\cal V}, \eea where
$E_\psi$ and $P_\psi$ are given by formulas (\ref{Et-psi}) and
(\ref{Pt-psi}) respectively. This means that the extra term ${\cal
V}$ play a role of a potential term.

The Friedmann equations of motion are
\begin{equation}
\begin{split}
3H^2&=\frac{1}{m_p^2}~\left(E_\psi+{\cal V}\right),
\\
3H^2+2\dot H&={}-\frac{1}{m_p^2}~\left( P_\psi-{\cal V} \right),
\end{split}
\label{eomprho-mod}
\end{equation}
Therefore
\begin{equation}
\dot H={}-\frac{1}{2m_p^2}\left(P_\psi+ E_\psi\right).
\label{dH-mod}
\end{equation}

 We choose such ${\cal V}$
 that $\psi_k$ in the non-flat case are the same as in
the flat case. Using (\ref{E-exp-psin}) and (\ref{P-exp-psin})) we
get

\begin{equation}
\label{Hdotgeneral} \dot
H=\frac{1}{2m_p^2}\sum_n\alpha_n^2\epsilon_n e^{\beta
\alpha_n^2}\Bigl(A^2_ne^{2\alpha_nt}+2A_nB_n+
B_n^2e^{-2\alpha_nt}\Bigr).
\end{equation}

Using (\ref{psi-n}) we can rewrite (\ref{Hdotgeneral}) as follows
\begin{equation}
\label{Hdotgeneral2} \dot H=\frac{1}{2m_p^2}\sum_n\epsilon_n
e^{\beta
\alpha_n^2}\dot\phi_n^2={}-\frac{1}{2m_p^2}\sum_n\left(2ce^{-2\alpha_n^2}+\xi^2\right)
\dot\phi_n^2.
\end{equation}

Substituting values of $\epsilon_n$ (formula \ref{eq-epsilon}) and
using formulas (\ref{Egen}) and (\ref{P-gen}), we obtain that
\begin{equation}
    \dot H=\frac{1}{2m_p^2}\left(\frac{\xi^2}{2}(\partial\phi)^2-c\int_{0}^{1}
    \left(\partial(e^{-\rho
\partial^2} \Phi)\right)\left(\partial(e^{ \rho \partial^2}
\Phi)\right) d \rho \right)=\frac{1}{2m_p^2}(E(\phi)+P(\phi)),
\end{equation}
where
\begin{equation}
\label{phi_sum}
    \phi(t)=\sum_n \phi_n(t)=\sum_n \left(
    A_ne^{\alpha_nt}+B_ne^{-\alpha_nt}
     \right).
\end{equation}

Therefore,
\begin{equation}
\label{Hgeneral}
    H(t)=\frac{1}{2m_p^2}\left(\sum_n 2A_nB_np_{\alpha_n}t
 -\sum_n \left(-\frac{A_n^2}{2\alpha_n}e^{-2\alpha_n
t}+\frac{B_n^2}{2\alpha_n}e^{2\alpha_n t}\right)p_{\alpha_n}\right)+H_0,
\end{equation}
where $H_0$ is an integration constant and we assume the sum goes
over the complex conjugated roots.

It is convenient to rewrite (\ref{Hdotgeneral2}) as follows
\begin{equation}
\dot
H={}-\frac{1}{2m_p^2}\sum_n\frac{p_{\alpha_n}}{\alpha_n^2}\dot\phi_n^2=
{}-\frac{1}{2m_p^2}\Bigl((\xi^2-2)\sum_n\dot\phi_n^2+2\xi^2\sum_n
\alpha_n^2\dot\phi_n^2\Bigr).\label{Hdotgeneral3}
\end{equation}

Thus to obtain the crossing of cosmological constant barrier one
should consider the case $\xi^2<2$ and the field $\phi(t)$, which
consists of at least two modes.  It is easy to see that $H(t)$ has
no singular point at finite time. For some values of parameters we
obtain bouncing solutions, which satisfy the conditions $H(0)=0$
and $\dot H(0)>0$.

 In the following subsections we construct effective potentials
for one-, two- and $N$-modes solutions. One-mode models can be
consider as toy-models, whereas two-fields models are more
realistic. Note that there are modern inflation models, for
example~\cite{Linde2fields,Buchbinder07}, which include two scalar
fields (see also~\cite{Linde07} and references therein). In  this
paper we describe a procedure to construct effective models and
analyse only the simplest properties of them. More detail analysis
and compare with the nonlocal model dynamic is a subject of our
future investigations.

\subsection{One Mode Solutions}

\subsubsection{One Real Root:  General case}
Let $\alpha$ is a real root of (\ref{5c}), without loss of
generality we can assume $\alpha>0$. The corresponding scalar
field is (see (\ref{psi-n})) as follows
\begin{equation}
    \psi(t)=Ae^{\alpha t} + B e^{-\alpha t}.
\end{equation}
The function $\psi(t)$ is a solution of the following first order
differential equation:
\begin{equation}
\label{de_psi}
    \dot\psi^2=\alpha^2\left(\psi^2-4AB\right).
\end{equation}

Using superpotential method~\cite{DeWolfe} (see
also~\cite{AKV,AKVtwofields}) we consider the Hubble parameter as
a function of $\psi$:
\begin{equation}
    H=W(\psi(t)) \qquad \Rightarrow \qquad \dot
    H=\frac{dW}{d\psi} \dot\psi.
\end{equation}
From (\ref{Hdotgeneral3}) we obtain
\begin{equation}
\frac{dW}{d\psi}={}-\frac{p_{\alpha}}{2m_p^2\alpha^2}\dot\phi=
{}\mp\frac{p_{\alpha}}{2m_p^2\alpha}\sqrt{\psi^2-4AB}
\end{equation}
and
\begin{equation}
    W(\psi)=\mp\frac{p_{\alpha}}{2m_p^2\alpha}\left(\frac{\psi}{2}\sqrt{\psi^2
    -4AB}-2AB\ln\left(\alpha\left(\psi
    +\sqrt{\psi^2-4AB}\right)\right)    \right)+H_0.
\end{equation}
The corresponding potential ${\cal V}(\psi)$ is
\begin{equation}
\label{Vonegen} {\cal
V}(\psi)=3m_p^2W^2-\frac{p_{\alpha}}{2\alpha^2}\dot\psi^2=
3m_p^2W^2-\frac{p_{\alpha}}{2}\left(\psi^2-4AB\right).
\end{equation}
We see that the potential depend on values of constants $A$ and $B$,
more exactly depend on value of the production $AB$,  and does not
depend on sign of $W(\psi)$. The potential ${\cal V}(\psi)$ is not a
polynomial, but in the limit of the flat space-time
($m_p^2\rightarrow\infty$) we obtain a quadratic polynomial. Note
that in the case of the flat space-time one can eliminate constants
$A$ and $B$ adding a constant to potential ${\cal V}(\psi)$, so in
this case the equations of motion do not depend on these constants.
In the following subsections we analyse in detail a few particular
cases.

\subsubsection{One Real Root: Decreasing Solution}
Let us consider a simplest particular case:

\begin{equation}
\label{sol-a}\psi=\exp(-mt),
\end{equation}
From (\ref{Hgeneral}) we have
\begin{equation}
H(t)=\frac{p_m}{4m_p^2m}e^{-2mt}+H_0. \label{psi1}
\end{equation}
Therefore to ensure that (\ref{sol-a}) and (\ref{psi1}) solve the
Friedmann equations we have to add to the action the following
potential
\begin{equation}
{\cal V}(\psi)=\frac{3}{16m_p^2m^2}p_m^2\psi^4
+\frac{3p_m}{2m}H_0\psi^2-\frac{p_m}{2}\psi^2+3m_p^2H_0^2.
 \label{V1-psi}
\end{equation}

Let us note that one gets the same potential for the unbounded
solution $\psi_1=\exp(mt)$.

The decreasing solution (\ref{sol-a}), (\ref{psi1}) corresponds to
the scale factor
\begin{equation}
a(t)=a_0\exp \left({}-\frac{p_m}{8m_p^2m^2}\exp(-2mt)+H_0t\right).
\label{cf-1}
\end{equation}
This solution has no singularities at finite $t$. It describes an
increasing Universe only in the case $p_m<0$. As we have seen in
Sect.\ref{pressure-real} the negative pressure is possible if
$c>1$ and $m^2<m_{max}^2$ that takes place if
$\xi^2\leq\xi_{max}^2$, where $\xi_{max}$ is given by
(\ref{xi-max}).

\subsubsection{One Real Root:  Odd  Solution}
Let us consider a odd solution
\begin{equation}
\label{sol-b}\psi=\sinh(mt).
\end{equation}
From (\ref{Hgeneral}) we have
\begin{equation}
H(t)={}-\frac{p_m}{8m^2_p m}\sinh (2mt) -\frac{p_m}{4m^2_p}t+H_0
 \label{a-odd-more}
\end{equation}
and the potential for $H_0=0$ is
\begin{equation}
{\cal V}(\psi)=\frac{p_m}{2}\left(-1+\frac{3 p_m}{8
m_p^2}\left(t+\frac{1}{2m}\sinh (2mt)\right)^2\right).
\end{equation}

 The potential as a function of $\psi_2$ is given by
\begin{equation}
{\cal V}(\psi)=-\frac{p_m}{2}+\frac{3 p^2_m}{16 m_p^2
m^2}\left[\psi^2(1+\psi^2) +2\psi\sqrt{1+\psi^2}\mbox{arcsinh}
(\psi)+\mbox{arcsinh}(\psi)^{2} \right].
\end{equation}

The odd increasing solution (\ref{sol-b}), (\ref{a-odd-more})
corresponds to the scale factor
\begin{equation}
a(t)=a_0\exp\left({}-\frac{p_m}{16m^2_p
m^2}\left(\cosh(2mt)+2m^2t^2\right)+H_0t\right). \label{a-odd-a}
\end{equation}
The scale factor (\ref{a-odd-a}) has no singularity. It increases
at large time if $p_m<0$.

\subsubsection{One Pure Imaginary Root}

Let us consider an odd solution
\begin{equation}
\label{sol-c-m} \psi =\sin (\mu t).
\end{equation}
From   (\ref{E-s}), (\ref{p-sin}) and (\ref{dH-mod}) we get
 explicitly the Hubble parameter
\begin{equation}
H(t)=\frac{ \pi_\mu}{4m^2_p}t+\frac{ \pi_\mu}{8\mu m^2_p}\sin (2\mu
t) +H_0. \label{H-sin}
\end{equation}

The potential for $H_0=0$ as a function of $\psi $ is given by
\begin{equation}
\label{Pot-4-psi} {\cal V}(\psi )=\frac12 \pi_\mu
+\frac{3\pi_\mu^2}{16 m^2_p\mu^2} \left[\psi ^2(1-\psi ^2)+2\psi
\sqrt{1-\psi ^2}\arcsin(\psi )+\arcsin(\psi )^2\right].
\end{equation}
Note, that this formula is valued only for $\psi ^2\leqslant 1$.
On this region  the potential (\ref{Pot-4-psi}) is convex and it
has an unique minimum.

The odd periodic  solution (\ref{sol-c-m}), (\ref{H-sin})
corresponds to the scale factor
\begin{equation}
a(t)=a_0\exp \left(-\frac{\pi_\mu}{16\mu^2m^2_p}\cos (2\mu
t)+\frac{\pi_\mu}{8m^2_p}t^2 +H_0t\right) \label{a-less}
\end{equation}
and we have an expansion with an acceleration if $\pi_\mu>0$.  In
this case at $H_0=0$ we obtain the bouncing solution.

\subsection{Solutions with the crossing of cosmological constant barrier}
\subsubsection{Pair of Complex Roots}

For the case of a complex root $\alpha=r+i\nu$ we consider the
following real solution
\begin{equation}
\psi_4 =\psi_\alpha +\psi_{ \alpha^*}, \qquad\mbox{where} \qquad
\psi_\alpha =e^{-\alpha t}, \qquad \psi_{\alpha^*}=e^{-\alpha^*t}.
\end{equation}

From (\ref{Hgeneral}) we have
\begin{equation}
H(t)=\frac{1}{4m_p^2}\left[\frac{p_\alpha}{\alpha}e^{-2\alpha t} +
\frac{p_{\alpha^*}}{\alpha^*}e^{-2\alpha^*
t}+H_0\right]=\frac{1}{4m_p^2}\left[\frac{p_\alpha}{\alpha}\psi_\alpha^2
+ \frac{p_{\alpha^*}}{\alpha^*}(\psi^{*}_{\alpha^*})^2+H_0\right].
\label{psi-psic}
\end{equation}
The corresponding potential is as follows
\begin{equation}
\begin{split}
{\cal
V}&=\frac{3}{16m^2_p}\Bigl[\alpha^2\left(\xi^2-2+2\xi^2\alpha^2\right)^2\psi_\alpha^4
+
\alpha^{*2}\left(\xi^2-2+2\xi^2\alpha^{*2}\right)\psi_{\alpha^*}^4+H_0^2+{}\\
{}&+2\alpha\alpha^*\left(\xi^2-2+2\xi^2\alpha^2\right)
\left(\xi^2-2+2\xi^2\alpha^{*2}\right)
\psi_\alpha^2\psi_{\alpha^*}^2+2H_0\left(\frac{p_\alpha}{\alpha}\psi_\alpha^2
+
\frac{p_{\alpha^*}}{\alpha^*}(\psi^{*}_{\alpha^*})^2\right)\Bigr].
\end{split}
\end{equation}

From (\ref{psi-psic}) it follows that
\begin{equation}
\label{dotH}
    \dot H={}-\frac{1}{2m_p^2}\left[p_\alpha e^{-2\alpha t} +
p_{\alpha^*}e^{-2\alpha^*
t}\right]={}-\frac{e^{-2rt}}{2m_p^2}\Bigl[
(p_\alpha+p_{\alpha^*})\cos(\nu
t)+i(p_\alpha-p_{\alpha^*})\sin(\nu t)\Bigr].
\end{equation}

It is easy to check that $p_{\alpha^*}=p_{\alpha}^*$, so
$q_1\equiv p_\alpha+p_{\alpha^*}$ and $q_2\equiv
i(p_\alpha-p_{\alpha^*})$ are real numbers. Formula (\ref{dotH})
can be rewrite in the following form:
\begin{equation}
\label{dotH2}
    \dot H={}-\frac{\sqrt{q_1^2+q_2^2}}{2m_p^2}e^{-2rt}
\sin(\nu t+\gamma),
\end{equation}
where $\gamma$ is defined by the following relations:
\begin{equation}
     \sin(\gamma)=\frac{q_1}{\sqrt{q_1^2+q_2^2}}, \qquad
     \cos(\gamma)=\frac{q_2}{\sqrt{q_1^2+q_2^2}}.
\end{equation}

As known the state parameter
\begin{equation}
\label{w} w=-1-\frac23\frac{\dot H}{H^2},
\end{equation}
so we obtain that $w$ crossing the barrier $w={}-1$ infinite
number of times.

For example, in the case: $\xi^2=0$ and $c=1$, when
$p_\alpha=-2\alpha^2$ and
\begin{equation}
\dot H=\frac{1}{m_p^2}\sum_n\dot\phi_n^2.
\end{equation}
If we choose $\alpha=\sqrt{\pi}+i\sqrt{\pi}$, then $p_\alpha=-2\pi
i$, $q_1=0$, $q_2=-4\pi$, $\gamma=\pi$  and
\begin{equation}
\dot H=\frac{1}{m_p^2}\left(\alpha^2e^{-2\alpha
t}+(\alpha^*)^2e^{-2\alpha^*
t}\right)={}-\frac{4\pi}{m_p^2}e^{-\sqrt{\pi}t}\sin(\sqrt{\pi}t-\pi).
\end{equation}

The Hubble parameter  (\ref{psi-psic}) corresponds to the
following scale factor
\begin{equation}
a(t)=a_0\exp\left\{H_0t-\frac{1}{8m_p^2}\left[\frac{p_\alpha}{\alpha^2}e^{-2\alpha
t} + \frac{p_{\alpha^*}}{\alpha^{*2}}e^{-2\alpha^*
t}\right]\right\}, \label{a-c}
\end{equation}
substituting  the explicit formula for $p_\alpha$ and
$p_{\alpha^*}$ we get
\begin{equation}
a(t)=a_0\exp\left\{H_0t-\left(\frac{(\xi ^2-2)}{4m_p^2}\cos(2\nu t)
-\frac{\xi ^2e^{-2rt}}{2m_p^2} \left[(r^2-\nu ^2)\cos(2\nu t) -2r\nu
\sin(2\nu t)\right]\right)\right\} \label{a-c-exp}.
\end{equation}
We see that a late time expansion regime corresponds only to
$H_0>0$ (compare with~\cite{AK}). Let us note that a constant part
of the Hubble parameter $H_0$ can be incorporated in a plane
solutions~\cite{AK}:
\begin{equation}
\psi_n(t)=A_ne^{\alpha_{n,H_0}t}+B_ne^{-\alpha_{n,H_0}t},
\end{equation}
where $\alpha _{n,H_0}$ is related with $\alpha _n$
as
\begin{equation}
\alpha_{n,H_0}^2+3H_0\alpha_{n,H_0}=\alpha _n^2.
\end{equation}
Using this approximations authors has been investigate the case of
the SFT inspired values of parameters $\xi^2$ and $c$ and obtain a
cosmic acceleration with a periodic crossing of the $w=-1$
barrier. In our paper we obtain the similar result for arbitrary
values of $\xi^2$ and $c$. Note that for some particular values of
these parameters we can obtain one-fold crossing of the $w=-1$
barrier.

\subsubsection{Two Real Roots Solutions}

The above-mentioned solutions for real roots $m_n$ correspond to
monotonic behaviour of the Hubble parameter. To describe
nonmonotonic behaviour let us consider the case of $c>1$ and
$\xi^2<\xi^2_{max}$. There exist two real roots of (\ref{5c})
$m_1$ and $m_2$ such that that $|m_1|<|m_2|$. The corresponding
solution to (\ref{1c}) is
\begin{equation}
\label{phi2exp}
  \psi=\psi_1+\psi_2,\quad \mbox{where}\quad \psi_1=Ae^{m_1t}, \quad \psi_2=Be^{m_2t},
\end{equation}
where $A$ and $B$ are constants. Without loss of generality we can
put $A=1$.

Using (\ref{Hdotgeneral}) we obtain
\begin{equation}
  \dot{H}={}-\frac{1}{2m_p^2}
  \Bigl(p_{m_1}e^{2m_1 t}+B^2p_{m_2}e^{2m_2
  t}\Bigr).
\end{equation}

Let us analyze a possibility $\dot{H}=0$, which correspond to
crossing of the cosmological constant barrier for the state
parameter $w$. In Subsection~\ref{pressure-real} we have obtained
that $p_{m_1}<0$ and $p_{m_2}>0$ (See Figs. \ref{xi2_m} and
\ref{p_m}). So, for any roots $m_1$ and $m_2$, there exist such real
$B$ that $\dot{H}=0$ at the point $t=t_1$:
\begin{equation}
    B={}\pm\sqrt{\frac{{}-p_{m_1}e^{2m_1 t_1}}{p_{m_2}e^{2m_2
  t_1}}}.
\end{equation}

We conclude that solutions (\ref{phi2exp}) correspond to
cosmological models with the crossing of $w={}-1$ barrier.

The Hubble parameter and the scale factor are as follows:
\begin{equation}
     H={}-\frac{1}{4m_p^2m_1m_2}
     \left(p_{m_1}m_2e^{2m_1t}+B^2p_{m_2}m_1e^{2m_2t}\right)+H_0,
\end{equation}
\begin{equation}
    a=a_0\exp\left(-\frac{1}{4m_p^2m_1^2m_2^2}
    \left(p_{m_1}m_2^2e^{2m_1t}+B^2p_{m_2}m_1^2e^{2m_2t}\right)+H_0t\right).
\end{equation}

If $m_1>0$ and $m_2<0$, then at late time
\begin{equation}
    \dot{H}\approx {}-\frac{1}{2m_p^2} p_{m_1}e^{2m_1 t}>0
\end{equation}
and
\begin{equation}
     H\approx{}-\frac{1}{4m_p^2m_1}p_{m_1}e^{2m_1t}>0.
\end{equation}

Using
\begin{equation}
 H=W(\psi_1,\psi_2)=\frac{{}-1}{4m_p^2m_1m_2}
     \left(p_{m_1}m_2\psi_1^2+p_{m_2}m_1\psi_2^2\right)+H_0,
\end{equation}
we obtain the fourth degree polynomial potential
\begin{equation}
{\cal V}(\psi_1,\psi_2)=3m_p^2\left(\frac{1}{4m_p^2m_1m_2}
     \left(p_{m_1}m_2\psi_1^2+p_{m_2}m_1\psi_2^2\right)-H_0\right)^2-\frac{1}{2}\left(p_{m_1}\psi_1^2+
     p_{m_2}\psi_2^2\right).
\end{equation}

We can conclude that all solutions (\ref{phi2exp}) correspond to
cosmological models with the crossing of $w={}-1$ barrier. Let us
remind in this context that models with a crossing of the $w=-1$
barrier are a subject of recent studies
\cite{Rubakov,Starobinsky,mukhanov,Andrianov,Odintsov,Nesseris}.
Simplest models include two scalar fields (one phantom and one
usual field, see \cite{AKVtwofields,Bo,Wei} and refs. therein). In
our case a nonlocality provides a crossing of the $w=-1$ barrier
in spite of the presence of only one scalar field. This fact has a
simple explanation. The crossing of the $w=-1$ in our case is
driven by an equivalence of our nonlocal model to a set of local
models some of which are ghosts.

\subsection{Cosmological models for N-mode solutions}
Let us construct a cosmological model, with $\dot H$, defined by
(\ref{Hgeneral}). To do this we use the superpotential
method~\cite{DeWolfe}. We consider the Hubble parameter as a
function (superpotential) of $\psi_n$:
$H(t)=W(\psi_1,\psi_2,\dots,\psi_N)$. To construct potential we use
the simplest form of the superpotential
\begin{equation}
    H(t)=W(\psi_1,\psi_2,\dots,\psi_N)=\sum_{n=1}^NW_n(\psi_n).
\end{equation}

Using formula
\begin{equation}
    \dot H=\sum_{n=1}^N\frac{d W_n}{d \psi_n}\dot\psi_n,
\end{equation}
we obtain, that (\ref{Hdotgeneral2}) is satisfied if
\begin{equation}
\frac{d W_n}{d
    \psi_n}={}-\frac{1}{2m_p^2}\sum_n\left(2ce^{-2\alpha_n^2}+\xi^2\right)
    \dot\psi_n.
\end{equation}
In the case of simple roots $\alpha_n$ the solutions
\begin{equation}
    \psi_n=A_ne^{\alpha_n t} + B_n e^{-\alpha_n t}
\end{equation}
satisfy the following first order differential equation:
\begin{equation}
\label{de_psi2}
    \dot\psi_n^2=\alpha_n^2\left(\psi_n^2-4A_nB_n\right).
\end{equation}

So,
\begin{equation}
 \frac{dW_n}{d\psi_n}={}-\frac{p_{\alpha_n}}{2m_p^2\alpha_n^2}\dot\psi_n=
{}\mp\frac{p_{\alpha_n}}{2m_p^2\alpha_n}\sqrt{\psi_n^2-4A_nB_n}
\end{equation}
and
\begin{equation}
    W_n(\psi_n)=\mp\frac{p_{\alpha_n}}{2m_p^2\alpha_n}\left(\frac{\psi_n}{2}\sqrt{\psi_n^2
    -4A_nB_n}-2A_nB_n\ln\left(\alpha(\psi_n
    +\sqrt{\psi_n^2-4A_nB_n})\right)    \right).
\end{equation}

The corresponding potential ${\cal V}(\psi_1,\dots,\psi_n)$ is
\begin{equation}
\label{Vgen} {\cal
V}(\psi_1,\dots,\psi_n)=3m_p^2W^2-\sum_n\frac{p_{\alpha_n}}{2\alpha_n^2}\dot\psi_n^2=
3m_p^2\left(\sum_nW_n\right)^2+\sum_n\frac{p_{\alpha_n}}{2}\left(\psi_n^2-4A_nB_n\right).
\end{equation}
We see that the potential and does not depend on sign of $W$. The
potential $V(\psi)$ is not a polynomial, but in the limit of the
flat space-time ($m_p^2\rightarrow\infty$) we obtain a quadratic
polynomial. Formula (\ref{Vgen}) is a straightforward generalization
of (\ref{Vonegen}). Note that the form of potential is not unique
(compare with~\cite{AKVtwofields}).

\section{Conclusions}

We have studied linear nonlocal models which violate the NEC. The
form of them is inspired by the SFT. These models have an infinite
number of higher derivative terms and are characterized by two
positive parameters, $\xi^2$ and $c$.

The  model with $c=1$ is a toy nonlocal model for the dilaton
coupling to the gravitation field. A distinguished feature of it is
the invariance under the shift of the dilaton field to a constant.

For  particular cases of the parameters $\xi^2$ and $c$ the
corresponding actions describe  linear approximations to the bosonic
\cite{Witten,ZW} and nonBPS fermionic \cite{NPB} cubic SFT as well
as  to the  nonpolynomial SFT \cite{Berkovich,BSZ}.

The case $\xi=0$ corresponds to a linear approximation to the p-adic
string~\cite{p-adic}. Let us note that recently a p-adics string
inflation model has been  considered \cite{Cline}.

In the flat case all solutions of the equation of motion are plane waves and are
controlled  by roots of the characteristic equation. Our
characteristic equation has complex distinctive simple roots. In
some particular cases there are single or double roots, which are real or
pure imaginary. The energy on plane waves is equal to zero
except for the cases of  couples of roots $(\alpha, -\alpha)$. The
pressure is a sum of one mode pressures. The pressure for the one
plane wave corresponding to a real root can be positive or negative
depending on parameters of the theory. For $c\leq 1$ the one mode
pressure is positive and for $c>1$ it could be negative or positive.

To study cosmological applications  we have investigated the
behaviour of the models in the Friedmann background. We have
performed this study  within an approximation scheme. A simplest
approximation is
 a local field approximation
 (or a mechanical analogous model in a terminology of
 \cite{AJ}). On an example of the free flat case we have shown that
 in special cases we can use a local two derivatives approximation, but
  the next derivative  approximations  exhibit  artifacts.
Followed \cite{AV}  we have used the Weierstrass product
representation to study finite mode approximations. As was noted in
\cite{AV} a straightforward application of the Ostrogradski method
to these approximations  indicates that energies are unbounded (an
eigenvalue problem
  for the unbounded hyperbolic Klein-Gordon equation on
manifolds is solved in \cite{KozVol}) and
 it is  expected
\cite{AV} that an incorporation of non-flat metric or nonlinear terms could
drastically change the situation.

 A  distinguished cosmological property of  these models
is  a crossing of
the phantom divide \cite{AK,K}. But there are also possibilities for other types of behaviour.
Namely, the toy dilaton model possesses
decreasing solutions describing asymptotical flat Universes (adding
a cosmological constant modifies  these solutions to
 near-de Sitter solutions). It also has   odd  bouncing solutions
describing  a contracting Universes meanwhile even  bouncing solutions are
forbidden.
 For special values of  parameters  corresponding to tachyon SFT models  there are  even bouncing solutions
with an accelerated expansion.

We have shown that for
some particular cases there are
deformations of the model such that  exact solutions of the linear problem are
inherited by nonlinear non-flat ones. This is similar to what was done before
for  local models \cite{AKV,0605229,0612026}.
A stability of  exact solutions in local models has been studied  in \cite{AKV}.
 We will study stability of our solutions in the future work.

%%%%%%%%%%%%%%%%%%%%%%%%%%%%%%%%%%%%%%%%
\section*{Acknowledgements}
This research is supported in part by RFBR grant 05-01-00758.
 The work of I.A. and L.J. is supported in part
by INTAS grant 03-51-6346  and Russian President's grant
NSh--672.2006.1. L.J. acknowledges the support of the Centre for
Theoretical Cosmology, in Cambridge. S.V. is  supported in part by
Russian President's grant NSh--8122.2006.2. I.A. and S.V. would like
to thank  A.S. Koshelev and I.V. Volovich for useful discussions.
L.J. would like to thank D. Mulryne for very helpful communications.

%%%%%%%%%%%%%%%%%%%%%%%%%%%%%%%%%%%%%%%%%
\appendix

\section{Calculations of Nonlocal  Energy Density and Pressure on Plane Waves}
In this appendix we calculate the energy density and pressure for
the following solution
\begin{equation}
    \phi=Ae^{\alpha_1 t}+Be^{\alpha_2 t},
\end{equation}
where $\alpha_1$ and $\alpha_2$ are different roots of (\ref{5c}).
We have
\begin{equation}
\label{E1}
    E(Ae^{\alpha_1 t}+Be^{\alpha_2 t})=E(Ae^{\alpha_1 t})+E(Be^{\alpha_2
    t})+E_{cross}(Ae^{\alpha_1 t},Be^{\alpha_2 t})=
    ABE_{cross}(e^{\alpha_1 t},e^{\alpha_2
    t}),
\end{equation}
where the functional $E_{cross}$ is defined as follows:
\begin{equation*}
E_{cross}(\phi_1,\phi_2)=\xi^2\partial\phi_1\partial\phi_2+\phi_1\phi_2+c\Phi_1\Phi_2+{}
\end{equation*}
\begin{equation*}
{}+c\int _0^1\left[\left(
e^{-\rho\partial^2}\Phi_1\right)\partial^2\left(
e^{-\rho\partial^2}\Phi_2\right)+\left(
e^{-\rho\partial^2}\Phi_2\right)\partial^2\left(
e^{-\rho\partial^2}\Phi_1\right)\right] d\rho-{}
\end{equation*}
\begin{equation*}
{}-c\int _0^1\left[\partial\left(
e^{-\rho\partial^2}\Phi_1\right)\partial\left(
e^{-\rho\partial^2}\Phi_2\right)+\partial\left(
e^{-\rho\partial^2}\Phi_2\right)\partial\left(
e^{-\rho\partial^2}\Phi_1\right)\right] d\rho,
\end{equation*}
\begin{equation*}
\Phi_1\equiv e^{{}-\partial^2}\phi_1,\qquad \Phi_2\equiv
e^{{}-\partial^2}\phi_2.
\end{equation*}
For $\phi_1=e^{\alpha_1 t}$ and $\phi_2=e^{\alpha_2 t}$ we have
\begin{equation*}
E_{cross}(e^{\alpha_1 t},e^{\alpha_2
t})=e^{(\alpha_1+\alpha_2)t}\Big\{
\alpha_1\alpha_2-1+ce^{({}-\alpha_1^2-\alpha_2^2)}+ {}
\end{equation*}
\begin{equation*}
    {}+ce^{-\alpha_1^2-\alpha_2^2}
    \int _0^1\left[\alpha_2^2e^{\rho(\alpha_2^2- \alpha_1^2)}
    +\alpha_1^2e^{\rho(\alpha_1^2- \alpha_2^2)}-\alpha_1\alpha_2
    \left(e^{\rho(\alpha_2^2- \alpha_1^2)}
    +e^{\rho(\alpha_1^2- \alpha_2^2)}\right)\right]d\rho\Big\}.
\end{equation*}
If $\alpha_1=-\alpha_2$, then it is easy to show that
\begin{equation}
E_{cross}(e^{\alpha_1 t},e^{-\alpha_1 t})=-2p_{\alpha_1}.
\end{equation}
where $p_\alpha$ is given by (\ref{pressure}).  We get
\begin{equation}
E(Ae^{\alpha_1 t}+Be^{-\alpha_1 t})= -2ABp_{\alpha_1}.
\end{equation}

In the opposite case ($\alpha_1\neq -\alpha_2$)
\begin{eqnarray}
E_{cross}(e^{\alpha_1 t},e^{\alpha_2 t})=E_{\alpha_1,\alpha_2}
e^{(\alpha_1+\alpha_2) t};
\end{eqnarray}
\begin{equation}E_{\alpha_1,\alpha_2}=
\alpha_1\alpha_2-1+\frac{ce^{({}-\alpha_1^2-\alpha_2^2)}\left(
\alpha_2^2e^{\alpha_2^2-\alpha_1^2}-
\alpha_1^2e^{\alpha_1^2-\alpha_2^2}-\alpha_1\alpha_2\left(
e^{\alpha_2^2-\alpha_1^2}-e^{\alpha_1^2-\alpha_2^2}\right)\right)
}{\alpha_2^2-\alpha_1^2}.
\end{equation}

Constants $\alpha_1$ and $\alpha_2$ are roots of (\ref{5c}),
therefore,
\begin{equation}
E_{\alpha_1,\alpha_2}=\frac{1}{\alpha_2^2-\alpha_1^2}
\Bigl\{\alpha_1\alpha_2\left(\left(\xi^2\alpha_2^2-ce^{-2\alpha_2}\right)-
\left(\xi^2\alpha_1^2-ce^{-2\alpha_1}\right)\right)+\alpha_2^2\xi^2\alpha_1^2
-\alpha_1^2\xi^2\alpha_2^2\Bigr\}=0.
\end{equation}

Note that, the equality $E_{cross}(e^{\alpha_1 t},e^{\alpha_2
t})=0$ at $\alpha_1\neq -\alpha_2$ also follows from the  energy
conservation low. Let us calculate the pressure $P(\phi)$ for the
solution $\phi=Ae^{\alpha_1 t}+Be^{\alpha_2 t}$.
\begin{equation}
    P\left(Ae^{\alpha_1 t}+Be^{\alpha_2 t}\right)=P
       \left(Ae^{\alpha_1 t}\right)+P\left(Be^{\alpha_2 t}\right)+
    P_{cross}\left(A e^{\alpha_1 t},B e^{\alpha_2 t}\right),
\end{equation}
where
\begin{equation}
\label{Pcross}
P_{cross}(\phi_1,\phi_2)=E_{k_{cross}}(\phi_1,\phi_2)+E_{nl2_{cross}}(\phi_1,\phi_2)
-E_{p_{cross}}(\phi_1,\phi_2)-E_{nl1_{cross}}(\phi_1,\phi_2),
\end{equation}
\begin{equation}
E_{k_{cross}}=\xi^2\partial\phi_1\partial\phi_2, \qquad
E_{p_{cross}}=\phi_1\phi_2+c\Phi_1\Phi_2,
\end{equation}
\begin{equation}
E_{nl1_{cross}}=c\int _0^1\left[\left(
e^{-\rho\partial^2}\Phi_1\right)\partial^2\left(
e^{-\rho\partial^2}\Phi_2\right)+\left(
e^{-\rho\partial^2}\Phi_2\right)\partial^2\left(
e^{-\rho\partial^2}\Phi_1\right)\right] d\rho,
\end{equation}
and
\begin{equation}
E_{nl2_{cross}}={}-c\int _0^1\left[\partial\left(
e^{-\rho\partial^2}\Phi_1\right)\partial\left(
e^{-\rho\partial^2}\Phi_2\right)+\partial\left(
e^{-\rho\partial^2}\Phi_2\right)\partial\left(
e^{-\rho\partial^2}\Phi_1\right)\right] d\rho.
\end{equation}

The functional
$E_{cross}=E_{k_{cross}}+E_{nl2_{cross}}+E_{p_{cross}}+E_{nl1_{cross}}$,
proving that $E_{cross}=0$, we obtain that
\begin{equation}
\label{Ekpnl_cross}
    E_{k_{cross}}+E_{nl2_{cross}}=0 \qquad \mbox{and}  \qquad
    E_{p_{cross}}+E_{nl1_{cross}}=0.
\end{equation}

Therefore
\begin{equation}
\label{Pcross0} P_{cross}\left(e^{\alpha_n t},
    e^{\alpha_k t}\right)=0, \quad\mbox{if}\quad \alpha_n\neq -\alpha_k.
\end{equation}

The straightforward calculations give that
\begin{equation}
\label{Pcross00}
P_{cross}\left(e^{\alpha_nt},e^{-\alpha_nt}\right)=0.
\end{equation}
So, $P_{cross}\left(e^{\alpha_nt},e^{\alpha_k t}\right)=0$ for all
$\alpha_n$ and $\alpha_k\neq\alpha_n$. The pressure is as follows:
\begin{equation}
    P\left(\sum_{n=1}^{2} C_n e^{\alpha_n t}\right)=\sum_{n=1}^{2}C_n^2P
    \left(e^{\alpha_n t}\right),
    \label{P2}
\end{equation}
where
\begin{equation}
\label{Poneexp} P\left(e^{\alpha t}\right)=p_{\alpha }e^{2\alpha t}.
\end{equation}

\end{document}